\documentclass[twocolumn]{aastex63}
\usepackage{graphicx,times}
\usepackage{amssymb,amsmath}
\usepackage{amsfonts}
\usepackage{natbib}
\shorttitle{The DTD of r- and s-Process Sources from Stellar Ages}
\shortauthors{Zhang \& Rix}
\usepackage{multirow}

\newcommand{\subfigure}[1]{#1}

\begin{document}
\title{Constraining the Delay Time Distribution of the r- and s-Process from Stellar Ages at Solar Metallicity}
\author{Meng Zhang}
\affiliation{Max-Planck Institute for Astronomy, K$\ddot{o}$nigstuhl 17, D-69117 Heidelberg, Germany }
\affiliation{National Astronomical Observatories, Chinese Academy of Sciences, Beijing 100101, People's Republic of China\\}
\author{Hans-Walter Rix}
\affiliation{Max-Planck Institute for Astronomy, K$\ddot{o}$nigstuhl 17, D-69117 Heidelberg, Germany }

\correspondingauthor{Meng Zhang}
\email{mezhang@mpia.de}

\begin{abstract}
We present a new approach to constraining the delay time distribution (DTD) governing r- and s-process element production. Rather than using metallicity as a proxy for time, we work directly with stellar ages at solar metallicity, where chemical equilibrium models are most applicable. We analyze two complementary samples: more than $10,000$ red clump (RC) thin disk stars at solar metallicity from LAMOST, and solar twins with high-precision abundances; both have well-determined stellar ages $\tau$. We fit chemical-equilibrium models with power-law DTDs to the evolution of [Eu/Mg]$(\tau)$ and [Ba/Mg]$(\tau)$ in each sample, taking europium as a canonical r-process element and barium as an s-process element, each referenced to magnesium as a near-instantaneous, CCSNe-dominated element. To validate the method, we first apply it to [Fe/Mg]$(\tau)$ and recover that $\approx\!70\%$ of Fe comes from a delayed channel with a declining DTD $\propto t^{-1.2}$, in agreement with well-established results for iron production in thermonuclear supernovae. It also shows that about $60\%$ of Eu and Ba forming today originate from strongly-delayed processes, and the inferred delayed-production rates of both Eu and Ba \emph{rise} with delay time, opposite to that of thermonuclear supernovae. These results provide direct temporal constraints on r- and s-process enrichment, complementary to metallicity-based chemical-evolution studies. Because we model abundance referenced to Mg, and the same trend is recovered across stars of different guiding-center radii, this appears unaffected by radial migration and chemical evolution varying with Galactocentric radius. If confirmed, it points to delayed neutron-capture production beyond NS--NS mergers.
\end{abstract}

\section{Introduction}
The astrophysical sites responsible for the rapid neutron-capture process (r-process), which produces approximately half of the elements heavier than iron remain incompletely understood. Although core-collapse supernovae (CCSNe) can produce r-process elements through various channels, including neutrino-driven winds and magneto-rotational explosions \citep{Kobayashi2020}, the gravitational wave detection of GW170817 and its associated kilonova provided direct evidence that neutron star mergers (NSM) are also significant r-process producers \citep{Abbott2017}. The key questions are: What fraction of r-process elements originates from delayed sources versus prompt sources? What is the delay time distribution (DTD) of the delayed sources, such as NSM, contributing the yields to galactic chemical evolution?

Substantial effort has been devoted to constraining r-process production from abundance patterns in metal-poor stars and chemical evolution modeling. Early works studied europium, a typical r-process element, which shows ``knee'' structures in the [Eu/Fe] versus [Fe/H] diagram similar to those of the $\alpha$-elements \citep{Cescutti2015, ShenSJ2015}. Following GW170817, \citet{Cote2019} demonstrated that reproducing the observed [Eu/Fe] decline at [Fe/H] $> -1$ in the Milky Way disk requires either a steeper-than-canonical DTD ($\propto t^{-1.5}$ rather than $t^{-1}$) or significant contributions from additional prompt channels. \citet{Simonetti2019} reached similar conclusions using stochastic chemical evolution models, requiring DTD $\propto t^{-1.5}$ to explain both the trend of [Eu/Fe] and the observed scatter at fixed metallicity. These studies converge on the finding that steep DTDs are necessary to explain the [Eu/Fe] evolution with metallicity, though debate continues about whether NSM are the exclusive r-process source or whether prompt channels contribute significantly at early times \citep{Skuladottir2020}.

However, all of these studies share a common methodological limitation: they use \emph{metallicity as a proxy for time} through chemical evolution models. The observed quantity is [Eu/Fe] versus [Fe/H], and the inference of DTD properties requires assumptions about the star formation history (SFH), gas inflows and outflows, stellar recycling, and the full network of element production. The ``knee'' in [Eu/Fe] versus [Fe/H] could arise from multiple effects, changes in the DTD, the onset of Type Ia supernovae producing iron, transitions between thick and thin disk populations, or variations in the SFH, making it challenging to isolate the DTD signature. Moreover, these approaches typically do not have access to individual stellar ages, instead relying on the simple assumption that lower-metallicity stars formed earlier.

Here we pursue a conceptually different approach enabled by recent advances in stellar age determination. We directly measure [Eu/Mg] as a function of stellar age $\tau_{\rm age}$ in stars at fixed solar metallicity ([Fe/H] $\approx$ 0.0 or [M/H] $\approx$ 0.0). This method has three key advantages: (1) By referencing europium to magnesium rather than iron, we create a differential measurement against the most CCSNe-dominated (or most instantaneous) element. (2) By using explicit stellar ages rather than chemical evolution tracks, we directly observe the temporal accumulation of delayed enrichment without requiring assumptions about the full metallicity evolution. (3) By restricting to solar metallicity, we work in a regime where chemical equilibrium models \citep{Weinberg2017} are most applicable, where stellar yields are least sensitive to metallicity, and where observational samples with high-quality ages are most abundant. Our approach is motivated by the empirical observation that [Eu/Mg] increases approximately linearly from $\sim$$-$0.1 dex in old stars ($\tau \sim 10$ Gyr) to $\sim$+0.1 dex in recent stars ($\tau \lesssim 2$ Gyr) among solar-metallicity disk stars, as seen in high-precision solar twin studies \citep{Spina2018, Bedell2018} and large spectroscopic surveys \citep[LAMOST;][]{Deng2012,Zhao2012,Liu2014}.

Such an age-tagged abundance approach to the r-process was previously taken by \citet{Tsujimoto2021}, who analyzed the same $79$ solar twins (i.e. at solar metallicity) and found the r-process-to-Fe ratio to vary systematically with age. \citet{Tsujimoto2021} attributed this variation largely to radial migration: at fixed (solar) metallicity, older stars were born at smaller Galactocentric radii, where enrichment proceeded faster, so that solar metallicity was reached earlier, and an age sequence at [Fe/H]$\,\approx 0$ is therefore partly a sequence in birth radius. However, that analysis left the DTD of iron and of r-processes tangled; this is a genuine concern for any age-based inference at fixed metallicity, and our analysis is designed to be less sensitive to it in two ways. First, we reference every abundance to Mg rather than Fe. Since $[{\rm Eu/Mg}] = [{\rm Eu/Fe}] - [{\rm Mg/Fe}]$, this removes by construction the $[{\rm Mg/Fe}]$ ($\alpha$-clock) term, which carries most of the known radial and age structure of the $[{\rm X/Fe}]$ ratios and is driven by the varying Fe (Type~Ia) contribution; what remains is the second-order radial gradient of $[{\rm Eu/Mg}]$ itself. Second, we exploit the LAMOST sample, for which guiding-center radii are available, to test directly whether the $[{\rm X/Mg}](\tau)$ trend persists across stars of different angular momentum, and thereby to quantify how much of the signal radial migration could plausibly explain (Section~\ref{sec:discussion}).

The power of this age-based approach lies in its ability to constrain the \emph{rate} at which delayed enrichment accumulates relative to the evolution of the star formation rate, quantified by $\alpha$ and $\beta$: the power-law index of the DTD and the exponential growth rate of the SFH, respectively. Combined with other observations, our method provides a direct temporal derivative of delayed enrichment. This makes it complementary to metallicity-based constraints and potentially less sensitive to systematic uncertainties in other aspects of chemical evolution modeling.

In this paper, we develop an analytic chemical-evolution framework to interpret [Eu/Mg] versus age measurements, building on the equilibrium-abundance formalism of \citet{Weinberg2017}. We show that the primary observable, the slope of [Eu/Mg] with age, directly constrains the differential accumulation of delayed versus prompt enrichment. Applying this model to solar-metallicity disk stars, we derive quantitative constraints on both the fraction of Eu originating from delayed sources and the shape of the DTD. Barium (Ba) is an s-process element, which is mainly synthesized in the interiors of asymptotic giant branch (AGB) stars \citep[e.g.,][]{Busso1999,Herwig2005}. In this work, we also use measurements of ‌[Ba/Mg] as a function of age‌ to constrain the DTD of the s-process. The paper is organized as follows: Section~\ref{sec:samples} introduces our two solar-metallicity samples; Section~\ref{sec:method} develops the chemical-equilibrium model, with parameterized forms for the SFH and DTD; Section~4 presents the fits to [Fe/Mg], [Eu/Mg], and [Ba/Mg], using Fe as a validation case; Section~\ref{sec:discussion} discusses the implications for r- and s-process nucleosynthesis; and Section~\ref{sec:conclusions} summarizes our conclusions. A detailed derivation of the equilibrium abundances is provided in the Appendix.

\section{Solar Metallicity Samples}
\label{sec:samples}
The elemental abundances of the interstellar medium enrich with cosmic time, and stars of different metallicities are born in different environments. In this work, we adopted solar metallicity stars as tracers for three key reasons. First, stars with solar-metallicity are more abundant than metal-poor stars in spectroscopic surveys. Second, restricting our study to a narrow range of metallicities effectively minimizes systematic uncertainties in elemental abundance measurements. Finally, stellar nucleosynthetic yields are known to vary as a function of metallicity. The yields of solar-metallicity stars have been characterized far more thoroughly and precisely than those of their metal-poor counterparts. We used two samples of stars with solar-metallicity: one is a red giant sample selected from LAMOST low-resolution spectroscopic survey, the other is the solar twins sample \citep[e.g.,][]{Bedell2014, Nissen2015, Spina2016a, Spina2016b} from high-resolution spectra.
\subsection{The LAMOST Red Clump Sample}
We use stellar parameters and abundances from LAMOST DR9 value-added catalog by \citet{ZhangM2025}. Using data-driven Payne ({\sc DD-Payne}), \citet{ZhangM2025} have derived 25 labels, including $T_{\rm eff}$, $\log\,g$, $v_{\rm mic}$, and abundances of 22 elements, namely C, N, O, Na, Mg, Al, Si, Ca, Ti, Cr, Mn, Fe, Ni, Ba, Sr, Y, Zr, La, Ce, Nd, Sm, and Eu, from the low-resolution spectra of LAMOST. The {\sc DD-Payne} is a hybrid approach that combines data-driven methods and physical priors for spectral modeling and fitting. As a data-driven approach, the stellar labels of the training set are taken from the APOGEE DR17 and GALAH DR3 high-resolution spectroscopic surveys; the neural-network model is then regularized during training with differential spectra from physical stellar-atmosphere models. For the elements central to this work, \citet{ZhangM2025} report typical internal precisions of $\approx\!0.03$--$0.05$\,dex for [Mg/Fe] and $\approx\!0.2$\,dex for the neutron-capture elements [Eu/Fe] and [Ba/Fe] at the high signal-to-noise of our RC sample.

We estimate spectroscopic ages for these giants with an XGBoost regression model \citep{Chen2016}, trained on $5{,}055$ APOKASC3 red giants \citep{Pinsonneault2025}, a mix of RGB and red-clump stars, whose asteroseismic ages are individually precise to better than $20\%$. The model uses four spectroscopic features: [Fe/H], [C/N] (formed as [C/Fe]$-$[N/Fe]), $T_{\rm eff}$, and $\log g$; the surface [C/N] ratio of a red giant encodes its main-sequence mass, and hence its age, after the first dredge-up. In $5$-fold cross-validation the model reaches an RMSE of $\approx\!2.1$\,Gyr and $R^2\approx0.6$; the fractional age error, measured from the scatter in $\ln(\tau_{\rm pred}/\tau_{\rm true})$, is $\approx\!0.3$--$0.4$ and mildly asymmetric (the $16$th--$84$th percentile range is $-0.23$ to $+0.37$), while the \emph{median} bias is negligible ($\approx\!2\%$). We then apply the trained model to the LAMOST DR9 red giants.

\begin{figure*}[htbp]
\centering
\includegraphics[width=0.97\textwidth]{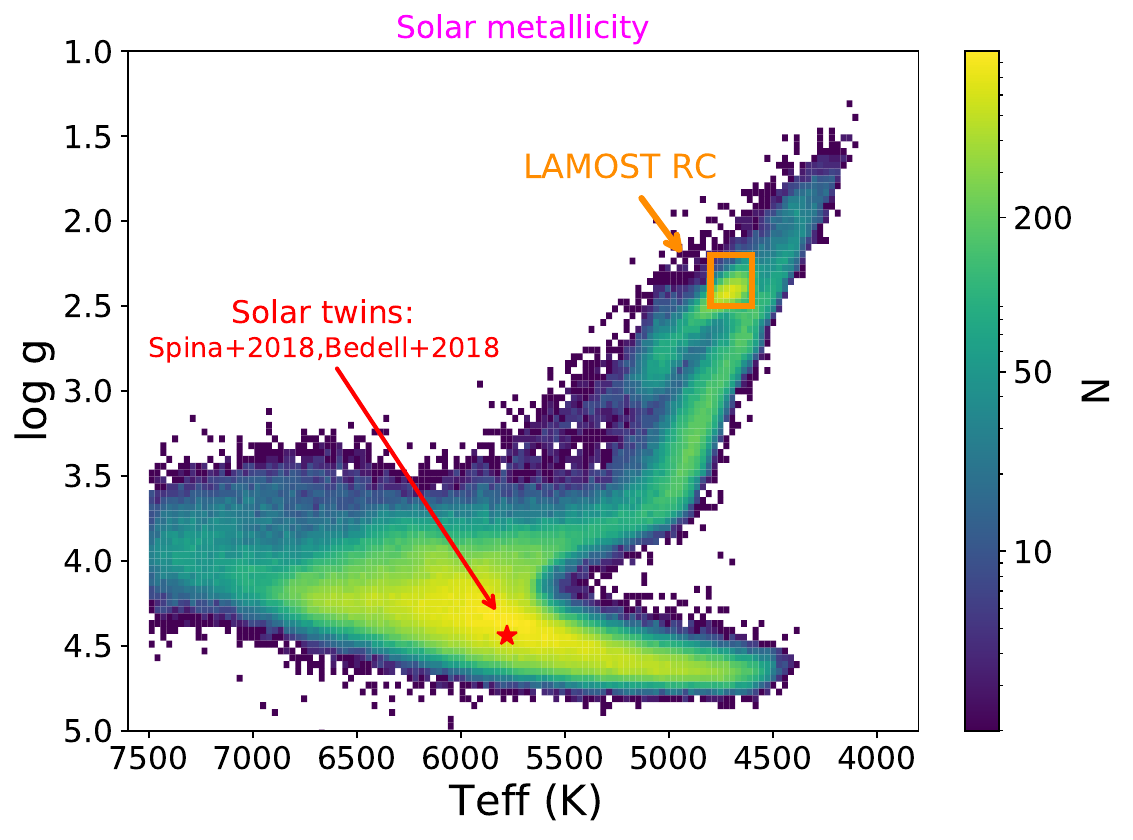}
\caption{The distribution of LAMOST sample stars at solar metallicity. The red clump stars in our sample are shown within the orange lines. The red star marks the location of the Sun.}
\label{fig:lmdr9_rc_select}
\end{figure*}

To minimize the systematic difference of our abundance measurements, we restricted our work to the red clump (RC) stars with very narrow parameter ranges around solar metallicity, predominantly belong to the low-$\alpha$ population.
Since the low-$\alpha$ disk stars typically have $J_{\rm{z}}\lesssim 25\,\rm{kpc km/s}$  \citep{TingYS2019, Gandhi2019}, whereas the high-$\alpha$ thick disk stars have larger $J_{\rm{z}}$ \citep{Gandhi2019}. To eliminate contamination of the high-$\alpha$ disk population, we exclude stars that have high-$J_{\rm{z}}$ or older than 9\,Gyr.
The sample stars are selected with:
\begin{itemize}
\item $4600 < T_{\rm eff} < 4800$\,K
\item $2.2 < \log g < 2.5$
\item $-0.1 < [{\rm Fe/H}] < +0.1$
\item $J_{\rm{z}} < 25$\, kpc km/s
\item $2.0 < \tau < 9.0$\,Gyr
\end{itemize}
Here we also exclude giants younger than 2.0\,Gyr, since the asteroseimic ages are less well calibrated for young stars, and the astroseismic-to-spectral training set is very small. The distribution of LAMOST RC sample stars are shown in Fig.~\ref{fig:lmdr9_rc_select}.

\begin{figure*}[htb!]
\centering
\subfigure
{
	\begin{minipage}{0.46\linewidth}
	\centering
	\includegraphics[width=0.95\columnwidth]{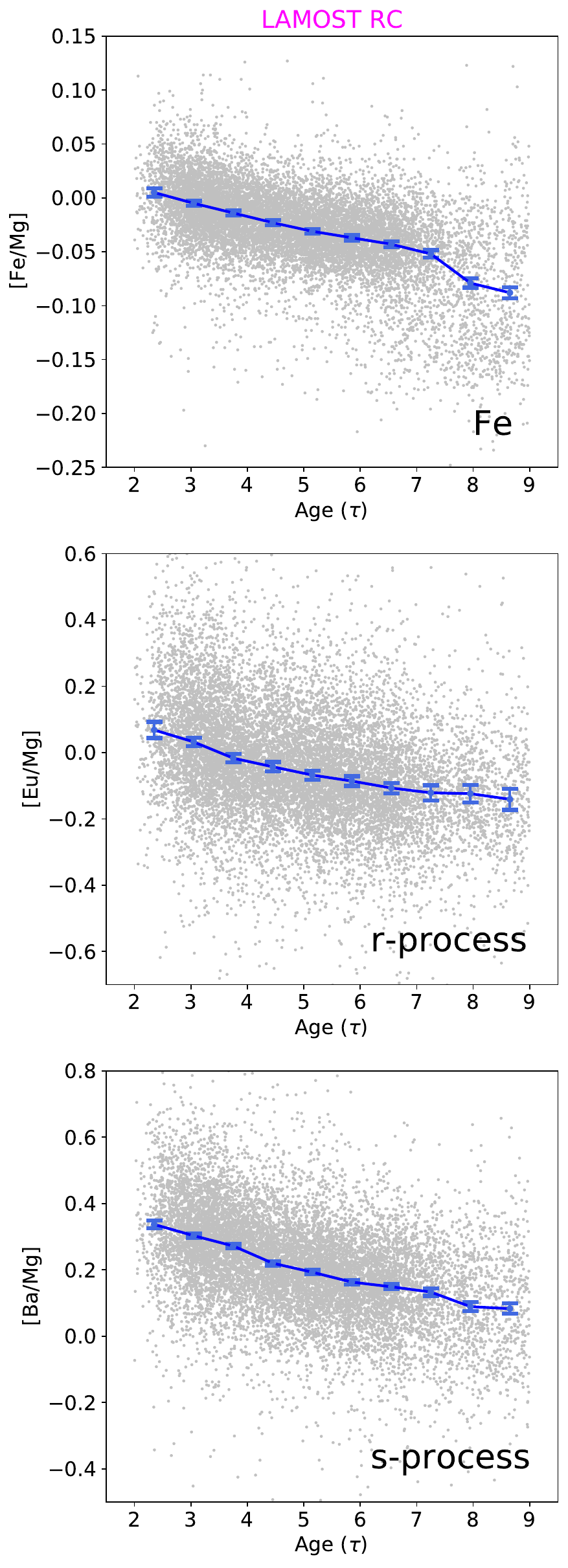}
	\end{minipage}
}
\subfigure
{
	\begin{minipage}{0.46\linewidth}
	\centering
	\includegraphics[width=0.95\columnwidth]{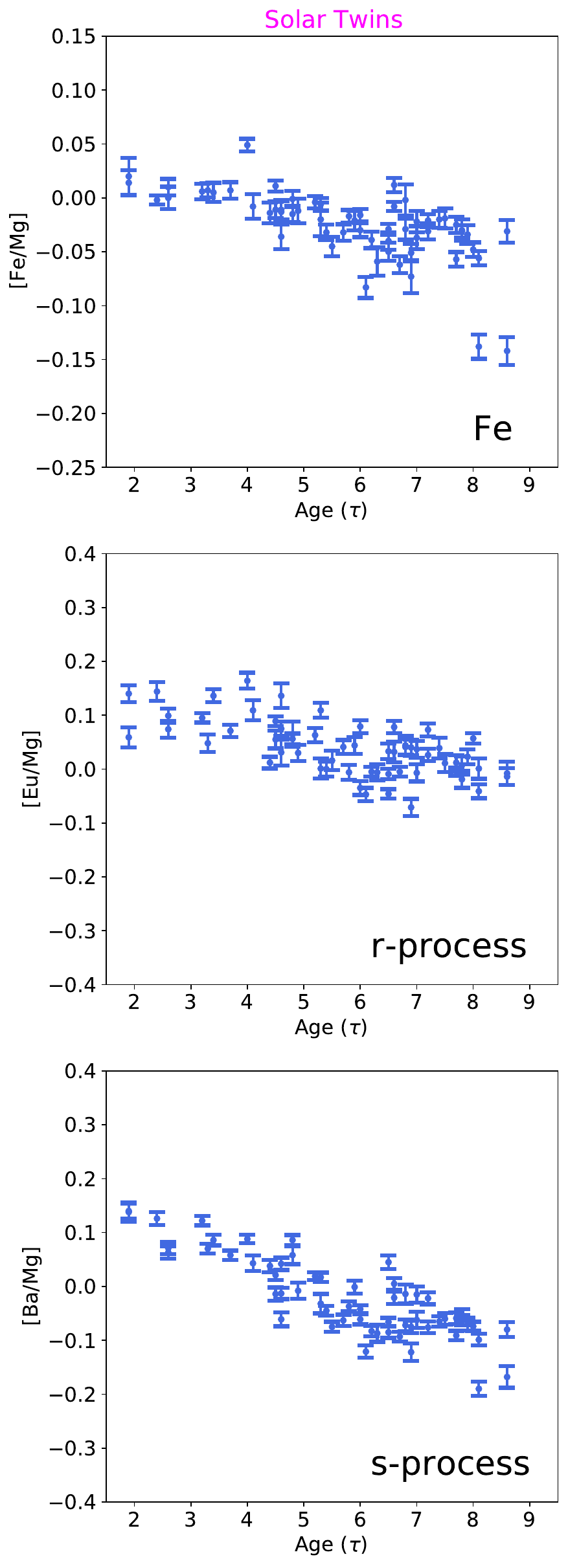}
	\end{minipage}
}
\caption{\textit{Left}: The blue dots show the LAMOST median values (and their uncertainties) for [Fe/Mg], [Eu/Mg], and [Ba/Mg] versus stellar age, all at [Fe/H]=0. The grey dots show the distribution of individual LAMOST sample stars. \textit{Right}: The blue dots show analogous [Fe/Mg], [Eu/Mg], and [Ba/Mg] measurements for solar twins (i.e. also at [Fe/H]$\sim 0$) versus stellar age.
\label{fig:data}}
\end{figure*}

Because individual ages carry $\sim\!30$--$40\%$ uncertainties, we do not fit individual stars; instead, we bin the sample in age and fit the binned medians of [X/Mg] (Fig.~\ref{fig:data}), the large sample size ($>\!10{,}000$ RC stars) keeping these median trends robust. 
 
Stars are separated into 10 age bins with each bin containing more than 300 stars. The blue dots represent the median [X/Mg] ratios of stars in each bin, with error bars indicating the uncertainties. The uncertainties are derived by dividing the median measurement errors by $\sqrt{N/10}$, where N is the number of stars in the respective bin, considering the typical systematic measurement errors of our sample.

 We note one systematic that the binning does \emph{not} remove: like any regression toward the training mean, the model mildly compresses the age scale, over-aging the youngest and under-aging the oldest stars, which can flatten the very [X/Mg]$(\tau)$ slopes we measure; 
this effect, however, does not appear to drive our results.

\subsection{The Solar-Twin Sample}
The solar-twin sample is drawn from \citet{Ramirez2014}: relative to the Sun, the sample stars have $T_{\rm eff}$ within $100$\,K and both $\log g$ and [Fe/H] within $0.1$\,dex. With this selection, $79$ solar twins with high-resolution, high signal-to-noise HARPS spectra were analyzed by \citet{Spina2018} and \citet{Bedell2018}. They used the same combined HARPS spectra and the same abundance-determination method for $30$ elements, including C, O, Na, Mg, Al, Si, S, Ca, Sc, Ti, V, Cr, Mn, Co, Ni, Fe, Cu, Zn, Sr, Y, Zr, Ba, La, Ce, Pr, Nd, Sm, Eu, Gd, and Dy. Because these are strictly differential measurements relative to the Sun, the abundances reach a precision of $\approx\!0.01$\,dex \citep{Spina2018, Bedell2018}, roughly an order of magnitude better than the LAMOST values, which is why the solar twins provide a valuable, if much smaller, cross-check. Using the isochrone method \citep[e.g.,][]{Vandenberg1985, Lachaume1999}, \citet{Spina2018} derived stellar ages with typical precisions of $0.4$\,Gyr.

In this work, we adopt stellar atmospheric parameters, including $T_{\rm eff}$, $\log\,g$, and elemental abundance measurements of [Fe/H], [Eu/H], [Ba/H] from \citet{Spina2018} and [Mg/H] from \citet{Bedell2018}. For consistency with the LAMOST RC sample, we further select solar twin stars whose median ages, with a 1$\sigma$ dispersion, fall within the identical 2-9\,Gyr range. Finally, there are 59 solar twins in our sample.

\section{Method}
\label{sec:method}
\subsection{Chemical Evolution Model}

We model the evolution of the Eu and Mg abundances for the low-$\alpha$ disk assuming chemical equilibrium, following \citet{Weinberg2017}; our differential, Mg-referenced treatment is also closely related in spirit to the empirical two-process decomposition of disk abundances into prompt and delayed contributions \citep{Weinberg2019, Griffith2019, Griffith2022}. Our approach parameterizes both the SFH and the DTD with simple functional forms that permit analytic solutions while capturing the essential physics.

We adopt the following assumptions:
\begin{enumerate}
\item \textbf{Chemical equilibrium:} The ISM abundance of each element adjusts instantaneously to variations in production and loss rates. We restricted stars to [Fe/H] $\approx$ 0 $\pm$ 0.1, where the Milky Way disk has reached chemical equilibrium \citep{Weinberg2017}. This equilibrium holds because the adjustment timescale ($\tau_{\rm adj} \sim M_{\rm gas}/[\psi(1+\eta-r)] \sim 0.5$--1.0\,Gyr at solar metallicity) is much shorter than the evolutionary timescales of the SFH and delayed enrichment accumulation ($\tau_{\rm evol} \sim 10$ Gyr). Here, $\eta$ denotes the loading factor of outflows and $r$ represents the loading factor of the gas recycling.

\item \textbf{Two enrichment channels:} Mg is produced only by prompt sources (CCSNe). Eu and Ba are produced by both prompt (CCSNe) and delayed sources (e.g., NSM).
\item \textbf{Functional forms for SFR and DTD:} We presume that in sufficient approximation the SFH follows 
\begin{equation}\psi(t) = \psi_0 e^{\beta t}
\end{equation}
and the delay time distribution follows
\begin{equation}
{\rm DTD}(\tau) = A\, \tau^{-\alpha}\ \ \ {\rm for}\ \  t_{\rm min} \leq \tau < t_{\rm max}.
\label{eq:DTDdef}
\end{equation}
\end{enumerate}
With this sign convention, $\alpha > 0$ corresponds to a DTD that \emph{declines} with delay time (as expected for, e.g., the canonical Type~Ia DTD, ${\rm DTD}\propto t^{-1}$), while $\alpha < 0$ corresponds to a DTD that \emph{increases} with delay time. Note that $\alpha$ is a dimensionless power-law index, whereas $\beta$ is a rate with units of Gyr$^{-1}$.

In this chemical equilibrium modeling \citep{Weinberg2017}, \emph{equilibrium} means that ISM abundances respond ``instantaneously" to the current balance of production and loss. But this balance can evolve over time.
At each epoch $t$, the ISM is in equilibrium with the production rate (including accumulated delayed enrichment) and the SFR at that time, and both quantities evolve over time. Stellar abundances of different ages record the evolution of the ISM composition at formation times.

The equilibrium assumption requires $\tau_{\rm adj} \ll \tau_{\rm evol}$. At solar metallicity, with $M_{\rm gas} \sim 10^{10}$ M$_\odot$, $\psi \sim 2$ M$_\odot$ yr$^{-1}$, and $(1+\eta-r) \sim 2.6$, we have $\tau_{\rm adj} \sim 0.5$--1.0\,Gyr. Meanwhile, the SFH e-folding time is $1/|\beta| \sim 10$\,Gyr, so the equilibrium condition is satisfied.

Under the equilibrium approximation (instantaneous adjustment to the current production/loss balance), the abundance ratio of Eu to Mg in the ISM at time $t$ is (see the Appendix for the full derivation):
\begin{equation}
\frac{Z_{\rm Eu}}{Z_{\rm Mg}} = \frac{y_{\rm Eu}^{\rm pr}}{y_{\rm Mg}^{\rm cc}}\left(1 + f \cdot \mathcal{R}(t)\right)
\end{equation}

where $y_{\rm Eu}^{\rm pr}$ and $y_{\rm Mg}^{\rm cc}$ are the CCSNe yields averaged by the initial mass function (IMF), $f \equiv y_{\rm Eu}^{\rm delay}/y_{\rm Eu}^{\rm pr}$ is the delayed to prompt yield ratio, and $\mathcal{R}(t)$ is the dimensionless \textbf{enrichment ratio}:
\begin{equation}
\mathcal{R}(t) = \int_{t_{\rm min}}^{t_{\rm eff}} \frac{\psi(t')}{\psi(t)} \cdot {\rm DTD}(t-t') \, dt'
\end{equation}

with $t_{\rm eff} = \min(t, t_{\rm max})$. This integral represents the accumulated delayed enrichment weighted by the SFH. The $\eta$ and $r$ cancel out of the ratio (see Appendix A).

For exponential SFR and power-law DTD, this integral evaluates analytically to:
\begin{equation}
\mathcal{R}(t) = A\int_{t_{\rm min}}^{t_{\rm eff}} e^{-\beta\tau} \cdot {\rm DTD}(\tau) \, d\tau
\end{equation}

where $A$ is the normalization constant that ensures $\int_{t_{\rm min}}^{t_{\rm max}}$ DTD($\tau$)$d\tau = 1$.

A star of age $\tau_{\rm age}$ was formed at time $t_{\rm form} = t_{\rm now} - \tau_{\rm age}$ and preserves the ISM composition at that epoch. The predicted abundance ratio is
\begin{equation}
\left[\frac{\rm Eu}{\rm Mg}\right](\tau_{\rm age}) = \left[\frac{\rm Eu}{\rm Mg}\right]_{\rm pr} + \log_{10}\left[1 + f \cdot \mathcal{R}(t_{\rm now} - \tau_{\rm age})\right],
\label{eq/model_prediction}
\end{equation}

where $[{\rm Eu/Mg}]_{\rm pr}$ sets the normalization for stars with only prompt enrichment. The key point is that the enrichment ratio $\mathcal{R}(t)$, and hence the slope of [Eu/Mg] with age, is sensitive to a \emph{combination} of the DTD index $\alpha$ and the SFH growth rate $\beta$: a greater increase in DTD with delay time can be partially compensated by a faster-growing SFH. This produces a strong $\alpha$--$\beta$ covariance in the posteriors (Fig.~\ref{fig:lmdr9_model}), so that the data constrain how rapidly delayed enrichment accumulates \emph{relative} to the evolution of the star-formation rate, but not the two rates entirely independently.

For physical interpretation, we define the \textit{delayed source fraction in recent stars}:
\begin{equation}
\begin{split}
f_{\rm delayed}({\rm today}) 
&\equiv \frac{Z_{\rm Eu}^{\rm delay}(t_{\rm now})}{Z_{\rm Eu}^{\rm pr}(t_{\rm now}) + Z_{\rm Eu}^{\rm delay}(t_{\rm now})}\\
&= \frac{f \cdot \mathcal{R}(t_{\rm now})}{1 + f \cdot \mathcal{R}(t_{\rm now})}
\end{split}
\end{equation}

This represents the fraction of Eu in stars that form today that originates from delayed sources (e.g., NSM). The complete model expressed in terms of this observationally motivated parameter is given in Eq.~(A15) of the Appendix.

The model has five free parameters:
\begin{itemize}
\item $\alpha$ (dimensionless): DTD power-law index (Eq.~\ref{eq:DTDdef})
\item $t_{\rm min}$ [Gyr]: minimum delay time of the delayed channel
\item $\beta$ [Gyr$^{-1}$]: SFH exponential growth rate
\item $[{\rm Eu/Mg}]_{\rm pr}$: prompt-only abundance normalization
\item $f_{\rm delayed}({\rm today})$: delayed fraction in recently-formed stars
\end{itemize}
with constraints $t_{\rm min} > 0$, and $0 < f_{\rm delayed}({\rm today}) < 1$. We fix $t_{\rm max} = t_{\rm now} = 13.8$ Gyr.

The primary degeneracy is between $\alpha$ and $\beta$: the slope of [Eu/Mg] versus age constrains $\alpha$ and $\beta$, while the amplitude constrains $f_{\rm delayed}({\rm today})$. Independent constraints on SFH from stellar age distributions or Gaia CMD modeling \citep{Ruiz-Lara2020} can help break this degeneracy. 

We fit the model to the data using Markov Chain Monte Carlo (MCMC) sampling with the \texttt{emcee} package \citep{Foreman-Mackey2013}. We adopt uniform priors on all parameters within physically motivated ranges and run 50 walkers for 4000 steps with a 2000-step burn-in. The likelihood assumes Gaussian measurement uncertainties.

\subsection{Model Predictions}
\label{sec:predictions}
\begin{figure*}[htbp]
\centering
\includegraphics[width=0.9\textwidth]{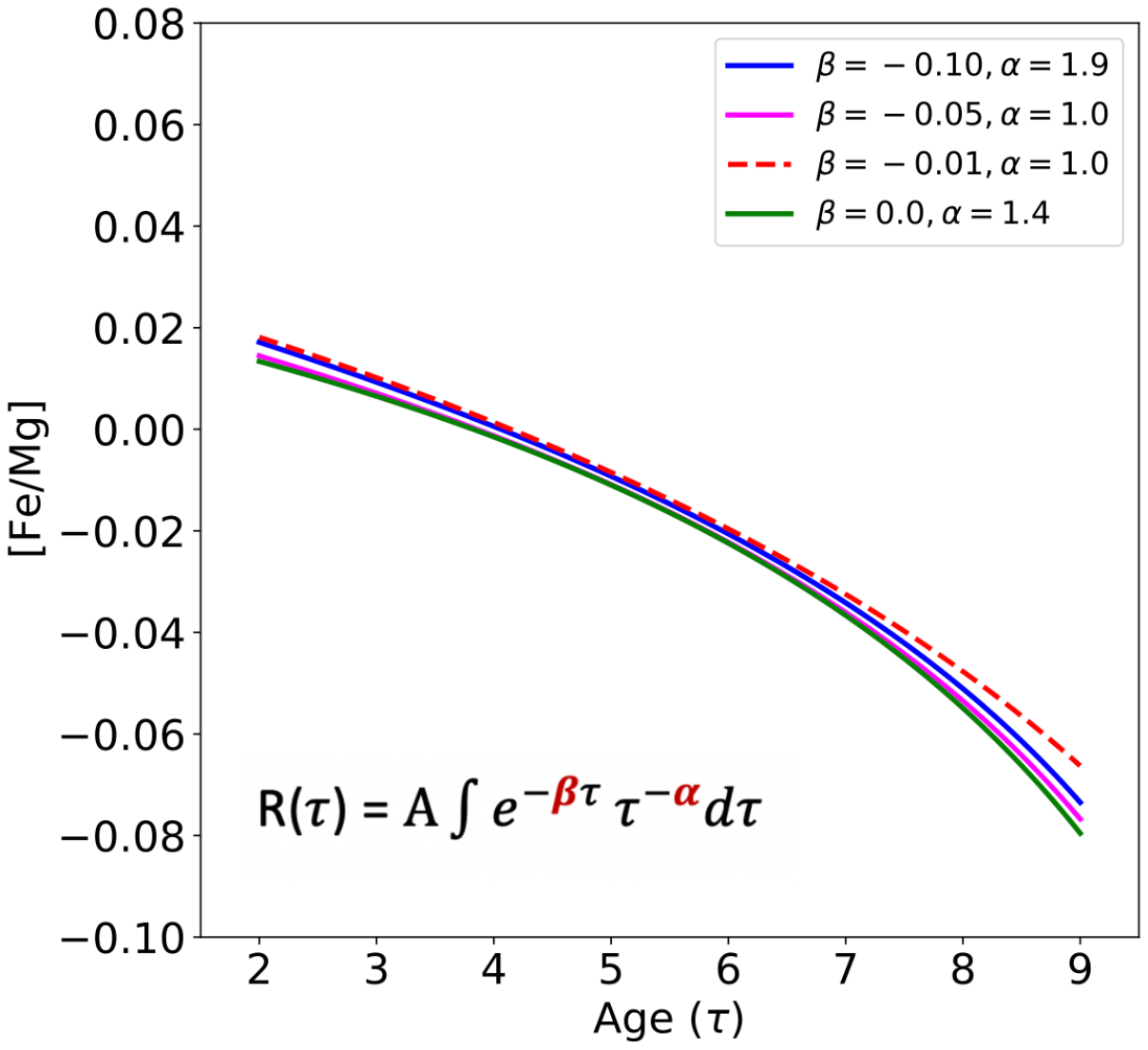}
\caption{Predictions for [Fe/Mg]$_{\rm [Fe/H]=0}$ from our model, for a range of star-formation histories (SFH growth rate $\beta$) and delay-time distributions (DTD index $\alpha$; see Eq.~\ref{eq:DTDdef}). The curves bracket the $(\alpha,\beta)$ values that fit the data best and illustrate the $\alpha$--$\beta$ degeneracy.}
\label{fig:lmdr9_model}
\end{figure*}

As an example, Figure~\ref{fig:lmdr9_model} shows model predictions of [Fe/Mg] for a range of DTD indices $\alpha$ and SFR $\beta$.A DTD that increases with delay time ($\alpha<0$) produces an [Fe/Mg] that climbs toward younger ages, because delayed enrichment continues to accumulate; the SFH rate $\beta$ modulates the amplitude and curvature of this climb. The near-degeneracy between $\alpha$ and $\beta$ is already visible here: several pairs $(\alpha,\beta)$ trace nearly the same curve over the $2$--$9$\,Gyr range that our data span.

\begin{figure*}[htb!]
\centering
\subfigure
{
	\begin{minipage}{0.48\linewidth}
	\centering
	\includegraphics[width=0.97\columnwidth]{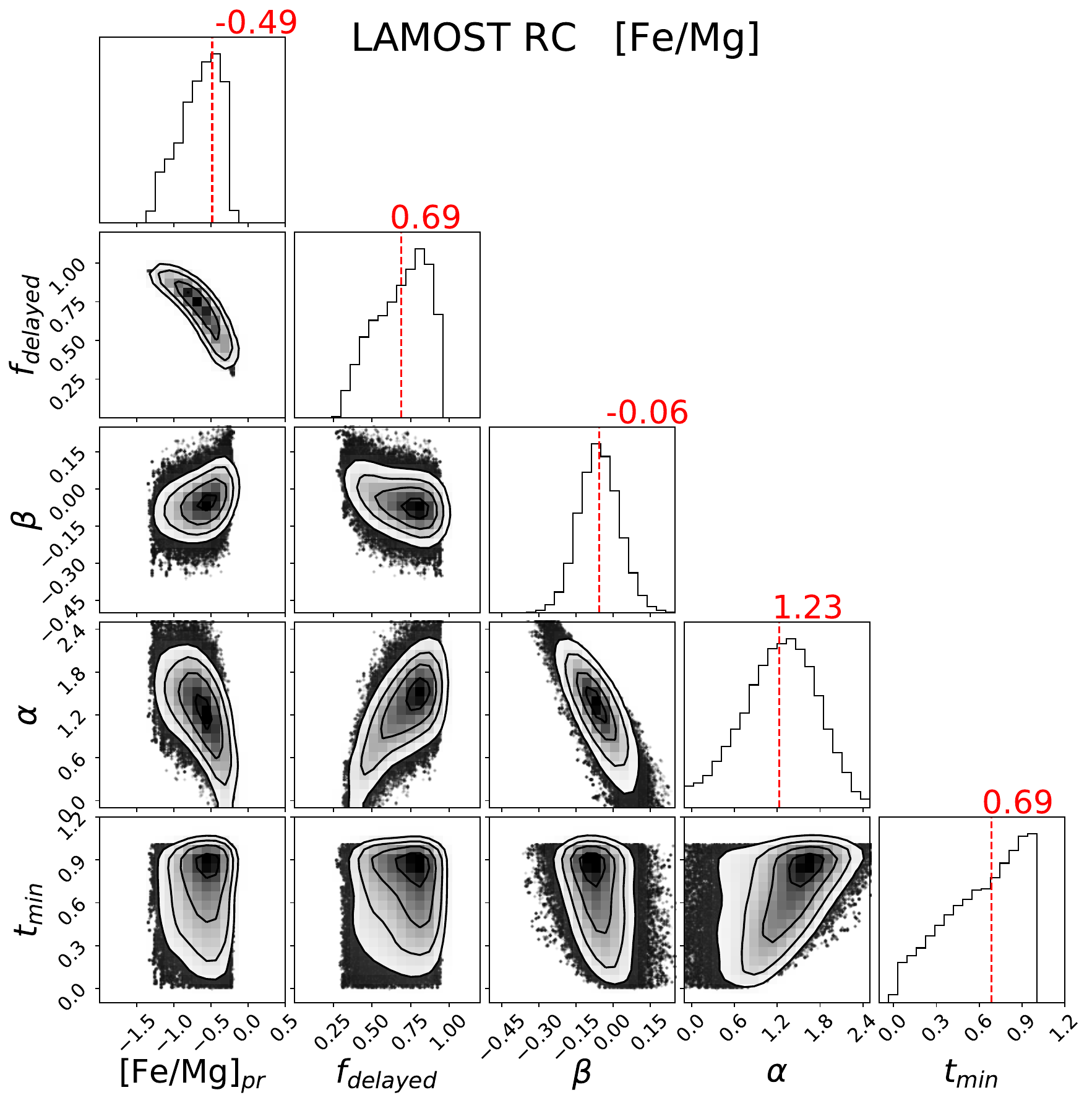}
	\end{minipage}
}
\subfigure
{
	\begin{minipage}{0.48\linewidth}
	\centering
	\includegraphics[width=0.97\columnwidth]{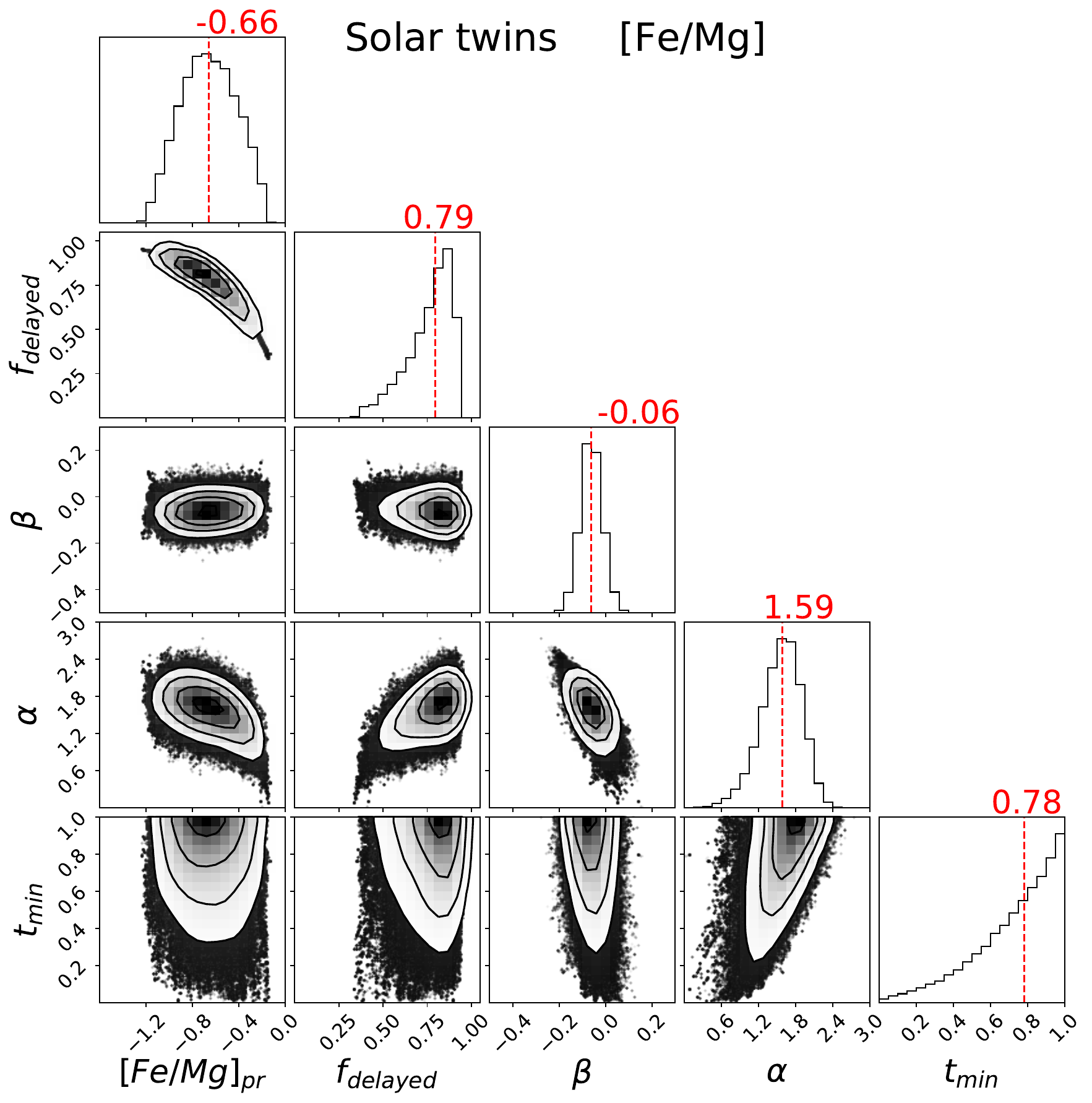}
	\end{minipage}
}
\caption{Posterior distributions for the [Fe/Mg] fit, for the LAMOST RC sample (left) and the solar twins (right). Diagonal panels show the 1D marginalized posteriors with their medians (red lines); off-diagonal panels show the 2D posteriors. The recovered DTD index is $\alpha \approx +1.2$ (LAMOST) and $\approx +1.6$ (twins), i.e.\ a \emph{declining} DTD, as expected for the Type~Ia channel that dominates Fe, with a delayed fraction $f_{\rm delayed} \approx 0.8$. Note the clear $\alpha$--$\beta$ covariance.
\label{fig:corner_fe}}
\end{figure*}

\begin{figure*}[htb!]
\centering
\subfigure
{
	\begin{minipage}{0.46\linewidth}
	\centering
	\includegraphics[width=0.95\columnwidth]{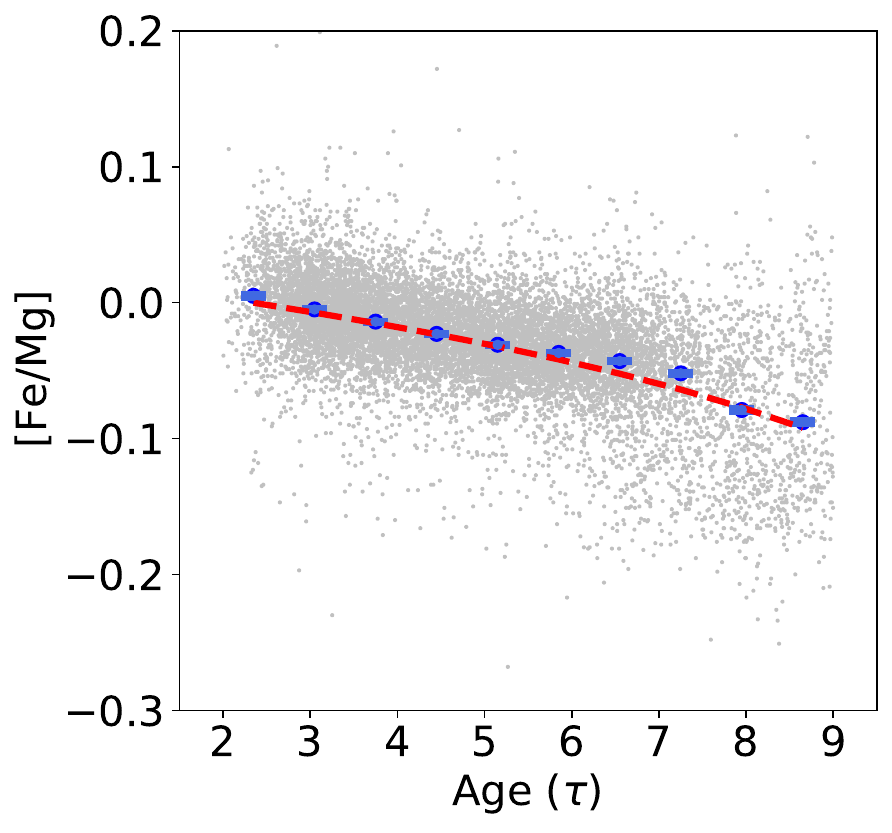}
	\end{minipage}
}
\subfigure
{
	\begin{minipage}{0.46\linewidth}
	\centering
	\includegraphics[width=0.95\columnwidth]{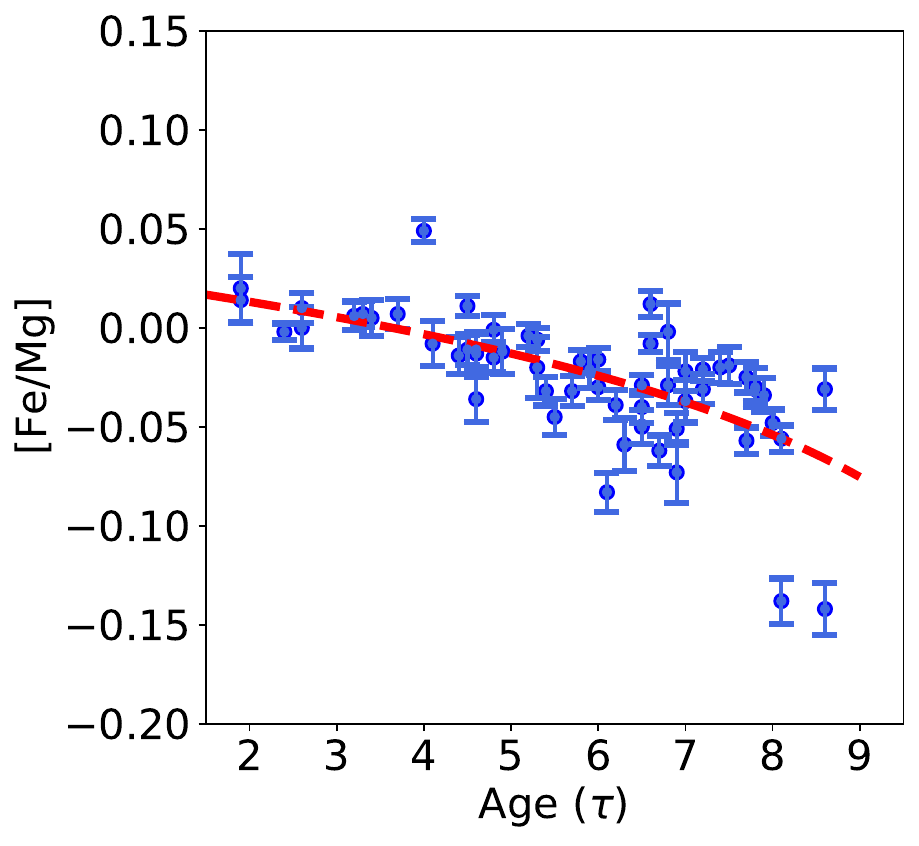}
	\end{minipage}
}
\caption{Chemical evolution model fits for [Fe/Mg]. \textit{Left}: The blue dots show the LAMOST median [Fe/Mg] versus stellar age, with error bars. The red dotted line shows the best-fit chemical model. The grey dots show the distribution of individual LAMOST sample stars. \textit{Right}: The blue dots show [Fe/Mg] for solar twins versus stellar age, with error bars.
\label{fig:results_fe}}
\end{figure*}

\section{Results}
\label{sec:results}
We fit each element ([Fe/Mg], [Eu/Mg], [Ba/Mg]) for LAMOST RC and solar twins samples, independently. The complete posteriors are shown in the contour figures, and all best-fit parameters are collected in Table~\ref{tab:fits}. We first describe the abundance trends that the model is asked to reproduce, then present iron as a validation case, and finally the r- and s-process elements Eu and Ba.

\subsection{Abundance Trends with Stellar Age}
\label{sec:trends}

Figure~\ref{fig:data} shows [Fe/Mg], [Eu/Mg], and [Ba/Mg] as a function of stellar age for both samples. All three ratios decline toward older ages, but with distinct shapes. [Fe/Mg] changes most rapidly among the oldest stars and flattens toward younger ones; [Eu/Mg], on the contrary, is nearly flat at old ages and rises noticeably for the recent stars ($\tau \lesssim 5$\,Gyr); [Ba/Mg] behaves similarly but with a steeper rise toward young ages than [Eu/Mg]. The solar twins show the same qualitative trends; for [Ba/Mg] there is a modest systematic offset between the two samples, plausibly reflecting the different abundance scales (Sec.~\ref{sec:samples}). These shapes already anticipate the fits: a sharply-declining (Type~Ia-like) DTD for Fe, and DTDs that rise with delay time for Eu and Ba.

\subsection{Method Validation: Iron}
\label{sec:fe}

Iron provides a stringent test of the method, because its enrichment history is comparatively well understood. About $70\%$ of the Fe in the solar neighborhood is thought to come from Type~Ia supernovae, the remainder from CCSNe, and the Type~Ia DTD is well established to be a declining power law $\propto t^{-1}$ over Gyr timescales \citep[e.g.,][]{Graur2015, Maoz2017,Friedmann2018, Wiseman2021}.
Applying our model to [Fe/Mg]($\tau$) (Figs.~\ref{fig:corner_fe} and \ref{fig:results_fe}, Table~\ref{tab:fits}), we recover a \emph{declining} DTD index $\alpha \approx +1.2$ for the LAMOST RC sample and $\approx +1.6$ for the solar twins, i.e.\ ${\rm DTD}\propto t^{-1.2}$ to $t^{-1.6}$, bracketing the canonical $t^{-1}$, and a delayed fraction $f_{\rm delayed}({\rm today}) \approx 0.70$ (LAMOST) and $\approx 0.80$ (solar twins). Both the DTD slope and the delayed fraction are in good agreement with independent determinations for Fe, which gives us confidence that the same machinery applied to Eu and Ba is meaningful.

\subsection{Europium: an r-Process Element}
\label{sec:eu}

\begin{figure*}[htb!]
\centering
\subfigure
{
	\begin{minipage}{0.48\linewidth}
	\centering
	\includegraphics[width=0.97\columnwidth]{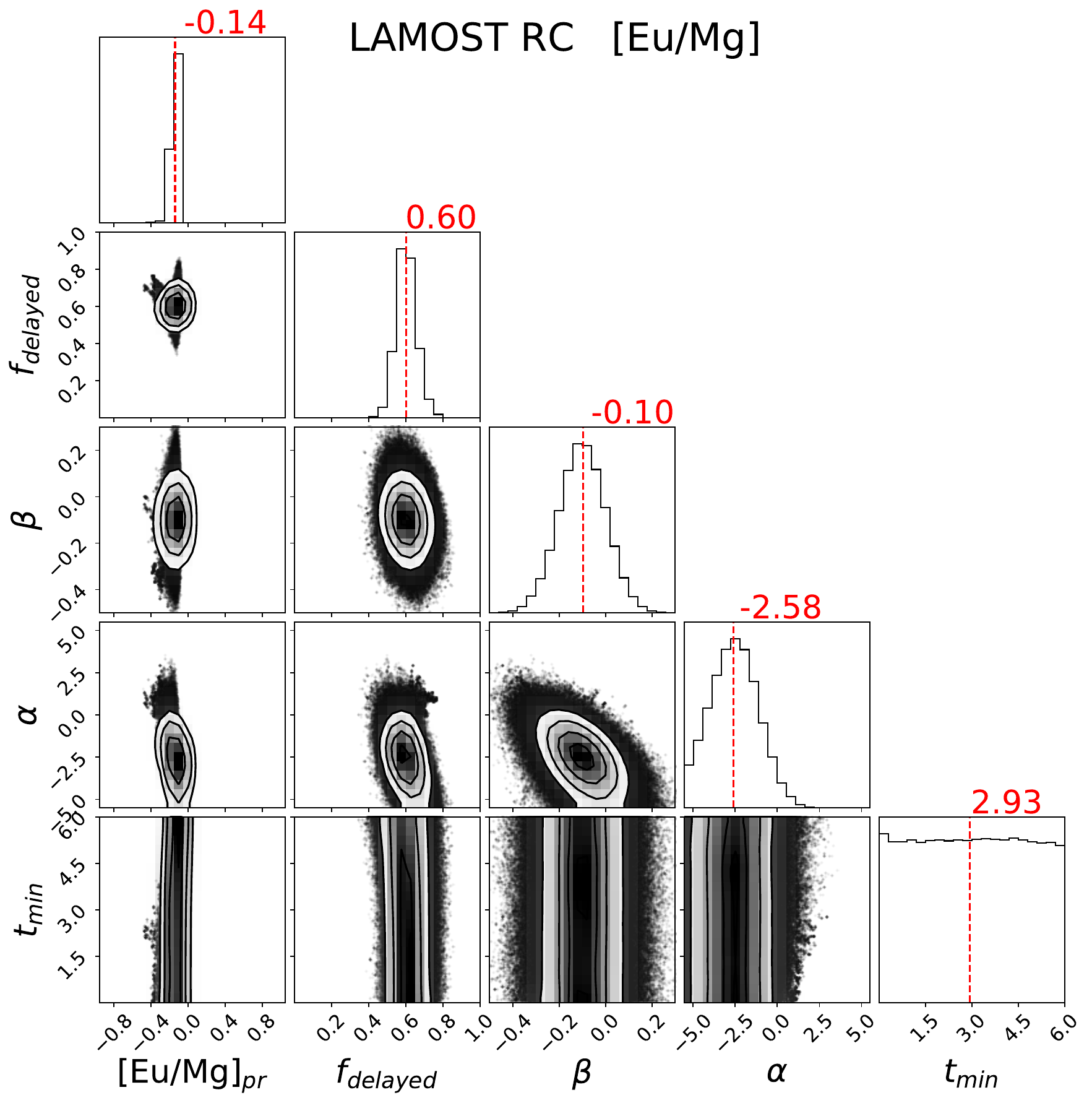}
	\end{minipage}
}
\subfigure
{
	\begin{minipage}{0.48\linewidth}
	\centering
	\includegraphics[width=0.97\columnwidth]{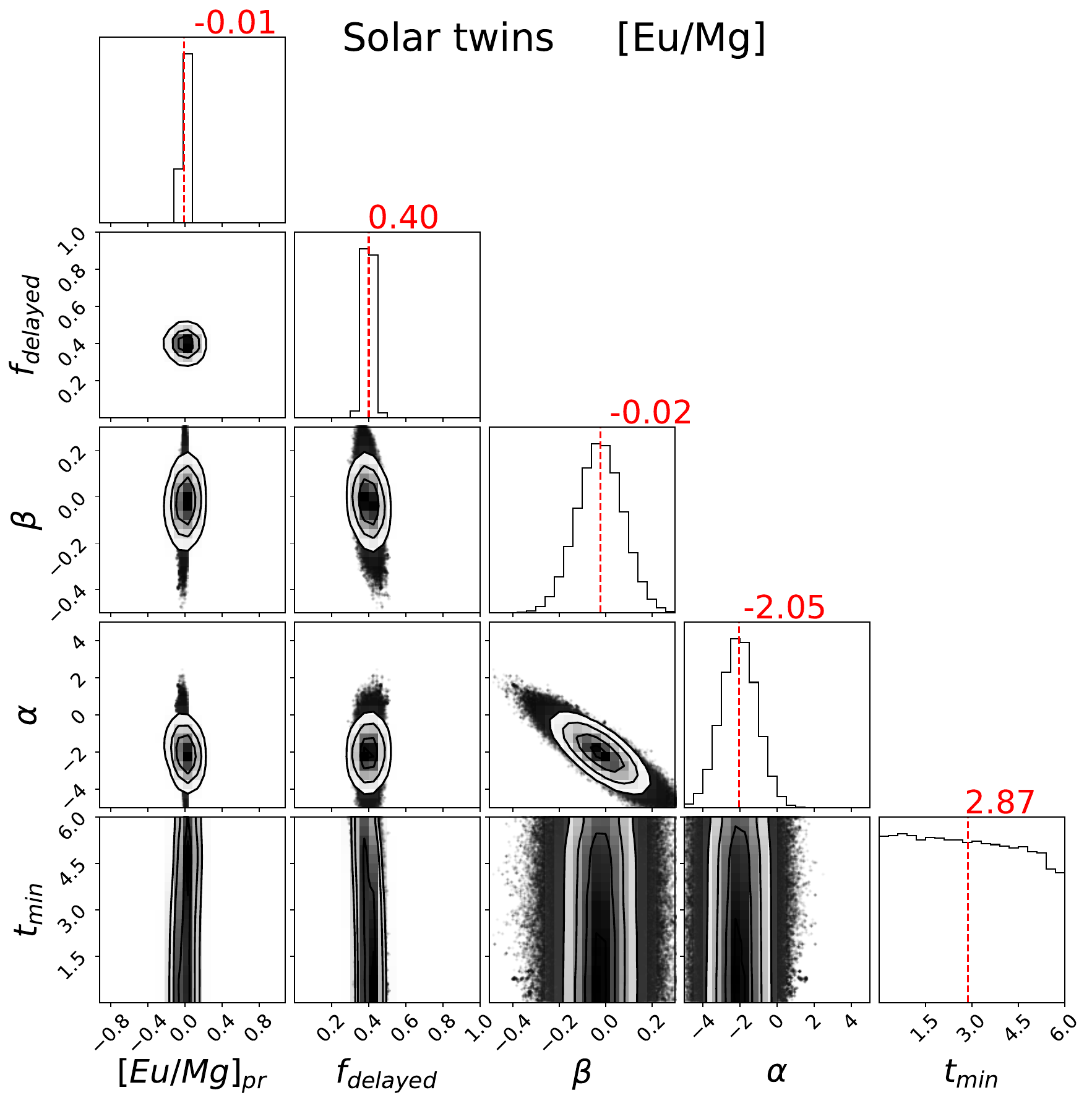}
	\end{minipage}
}
\caption{Posterior distributions for the [Eu/Mg] fit, for the LAMOST RC sample (left) and the solar twins (right). Diagonal panels show the 1D marginalized posteriors with their medians (red lines); off-diagonal panels show the 2D posteriors. The recovered DTD index is $\alpha \approx -2.8$ (LAMOST) and $\approx -2.2$ (twins), i.e.\ a DTD that \emph{rises} steeply with delay time, with a delayed fraction $f_{\rm delayed}\approx 0.6$ (LAMOST) and $\approx 0.45$ (twins). As for all elements, $\alpha$ and $\beta$ are covariant, and $t_{\rm min}$ is only weakly constrained.
\label{fig:corner_eu}}
\end{figure*}

Figure~\ref{fig:corner_eu} shows the posteriors for the [Eu/Mg] fit to the LAMOST RC sample. The best-fit parameters (medians with 16th/84th percentiles) are
\begin{align}
\alpha &= -2.58^{+1.45}_{-1.61} \\
t_{\rm min} &= 2.9^{+2.0}_{-2.0} \text{ Gyr} \\
\beta &= -0.1^{+0.1}_{-0.1} \text{ Gyr}^{-1} \\
[{\rm Eu/Mg}]_{\rm pr} &= -0.18^{+0.02}_{-0.03}  \\
f_{\rm delayed}({\rm today}) &= 0.60^{+0.06}_{-0.05} .
\end{align}
The negative index, $\alpha \approx -2.58$, corresponds to a DTD that \emph{rises} steeply with delay time (${\rm DTD}\propto \tau^{2.58}$), the opposite behavior to iron. The solar-twin sample gives a consistent, if shallower and noisier, result ($\alpha \approx -2.05$, $f_{\rm delayed}\approx 0.4$; Table~\ref{tab:fits}). The minimum delay time $t_{\rm min}$ is only weakly constrained in both samples (its posterior is nearly flat up to $\sim\!3$\,Gyr; Fig.~\ref{fig:corner_eu}), so the value above should be read as an indicative scale rather than a precise measurement.

\begin{figure*}[htb!]
\centering
\subfigure
{
	\begin{minipage}{0.46\linewidth}
	\centering
	\includegraphics[width=0.95\columnwidth]{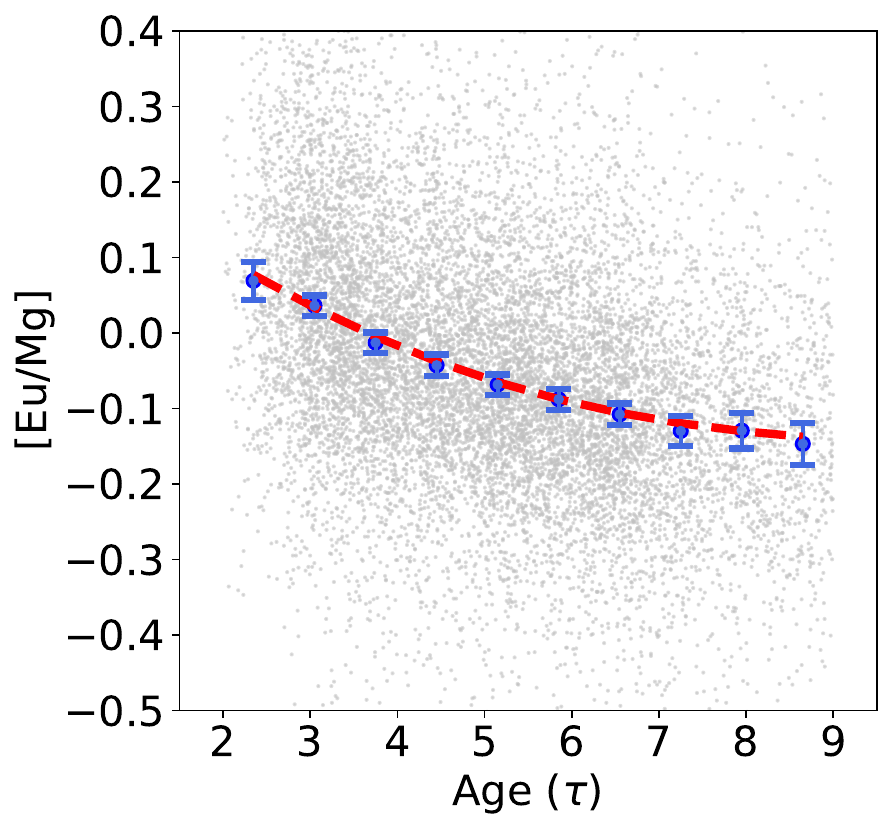}
	\end{minipage}
}
\subfigure
{
	\begin{minipage}{0.46\linewidth}
	\centering
	\includegraphics[width=0.95\columnwidth]{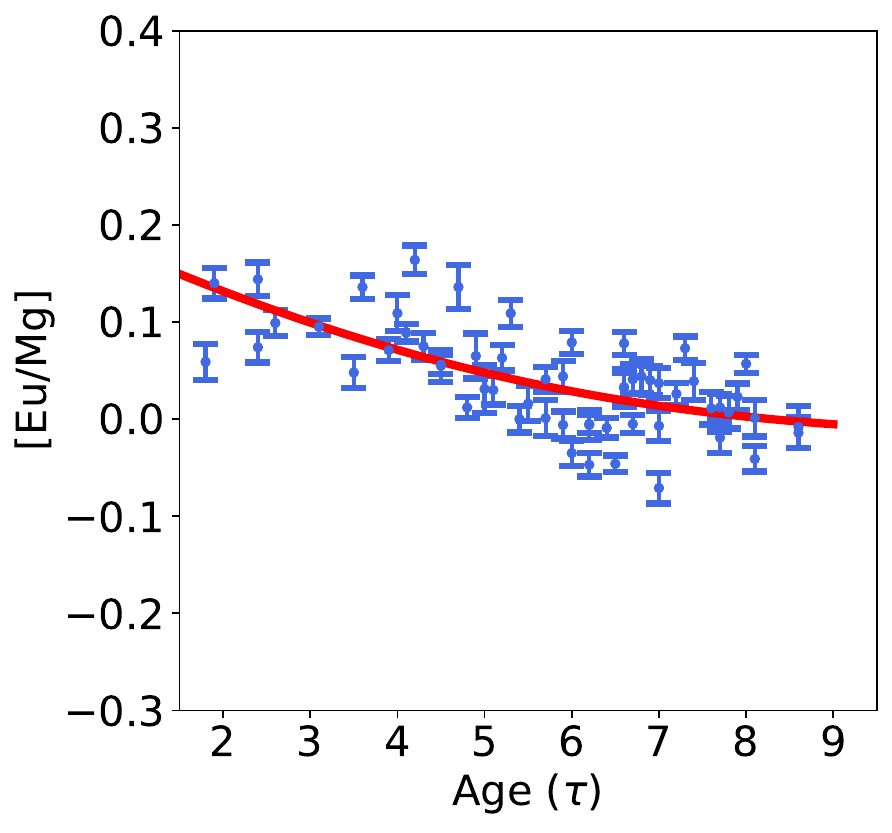}
	\end{minipage}
}
\caption{Chemical evolution model fits for [Eu/Mg]. \textit{Left}: The blue dots show the LAMOST median [Eu/Mg] versus stellar age, with error bars. The red dotted line shows the best-fit chemical model. The grey dots show the distribution of the $\sim$10,000 individual LAMOST sample stars. \textit{Right}: The blue dots show [Eu/Mg] for solar twins versus stellar age, with error bars. The red line shows the best-fit chemical model for these solar twin stars.
\label{fig:results_eu}}
\end{figure*}

Figure~\ref{fig:results_eu} shows the corresponding fit in the data plane. The median model (red line) passes through the binned [Eu/Mg] points within the uncertainties. The right panel shows that that the results of the best-fit model is consistent with the observational data with error bars, so the rising-DTD solution is a good description of the data and not over-fit.

\subsection{Barium: an s-Process Element}
\label{sec:ba}

\begin{figure*}[htb!]
\centering
\subfigure
{
	\begin{minipage}{0.48\linewidth}
	\centering
	\includegraphics[width=0.97\columnwidth]{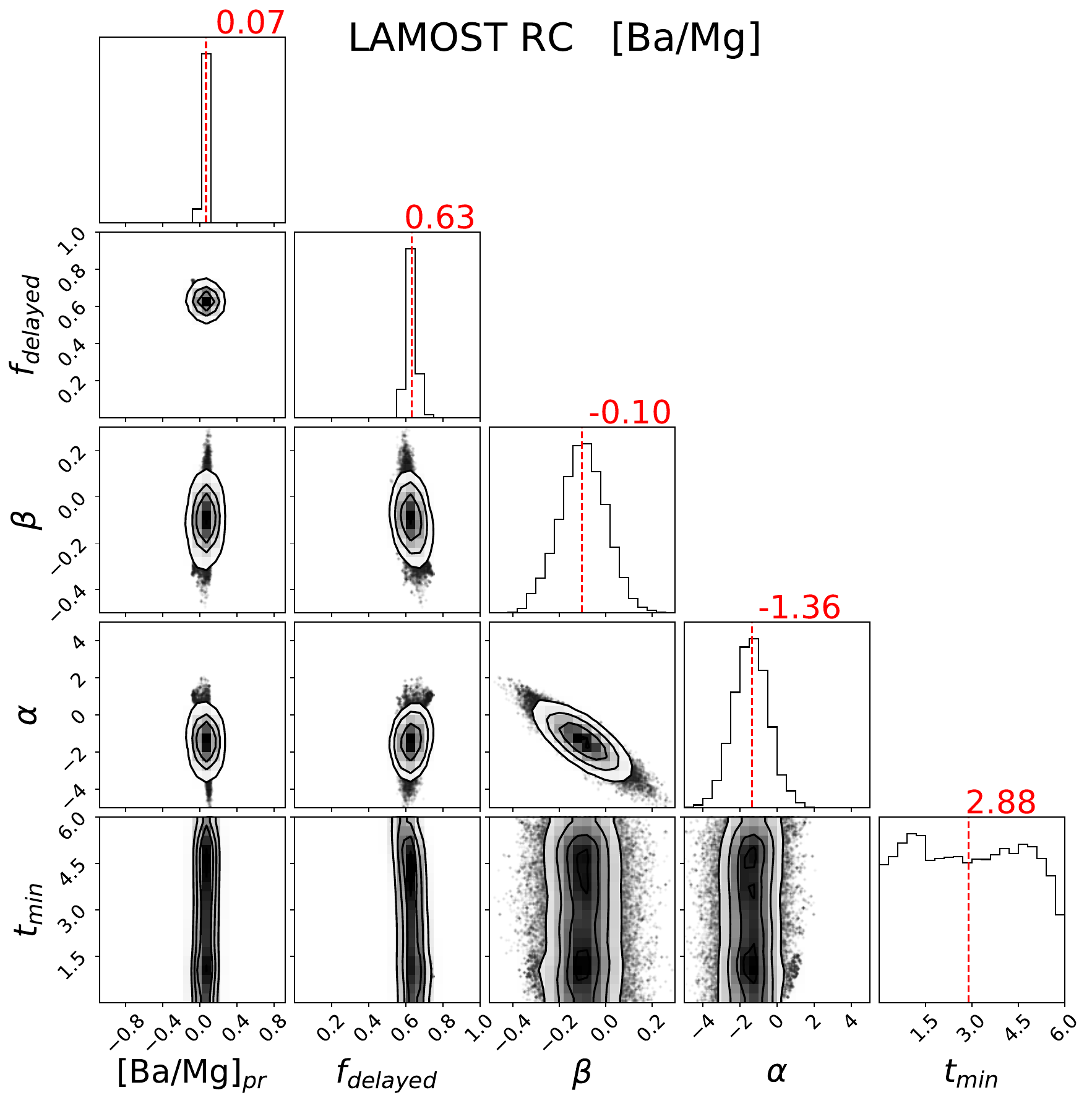}
	\end{minipage}
}
\subfigure
{
	\begin{minipage}{0.48\linewidth}
	\centering
	\includegraphics[width=0.97\columnwidth]{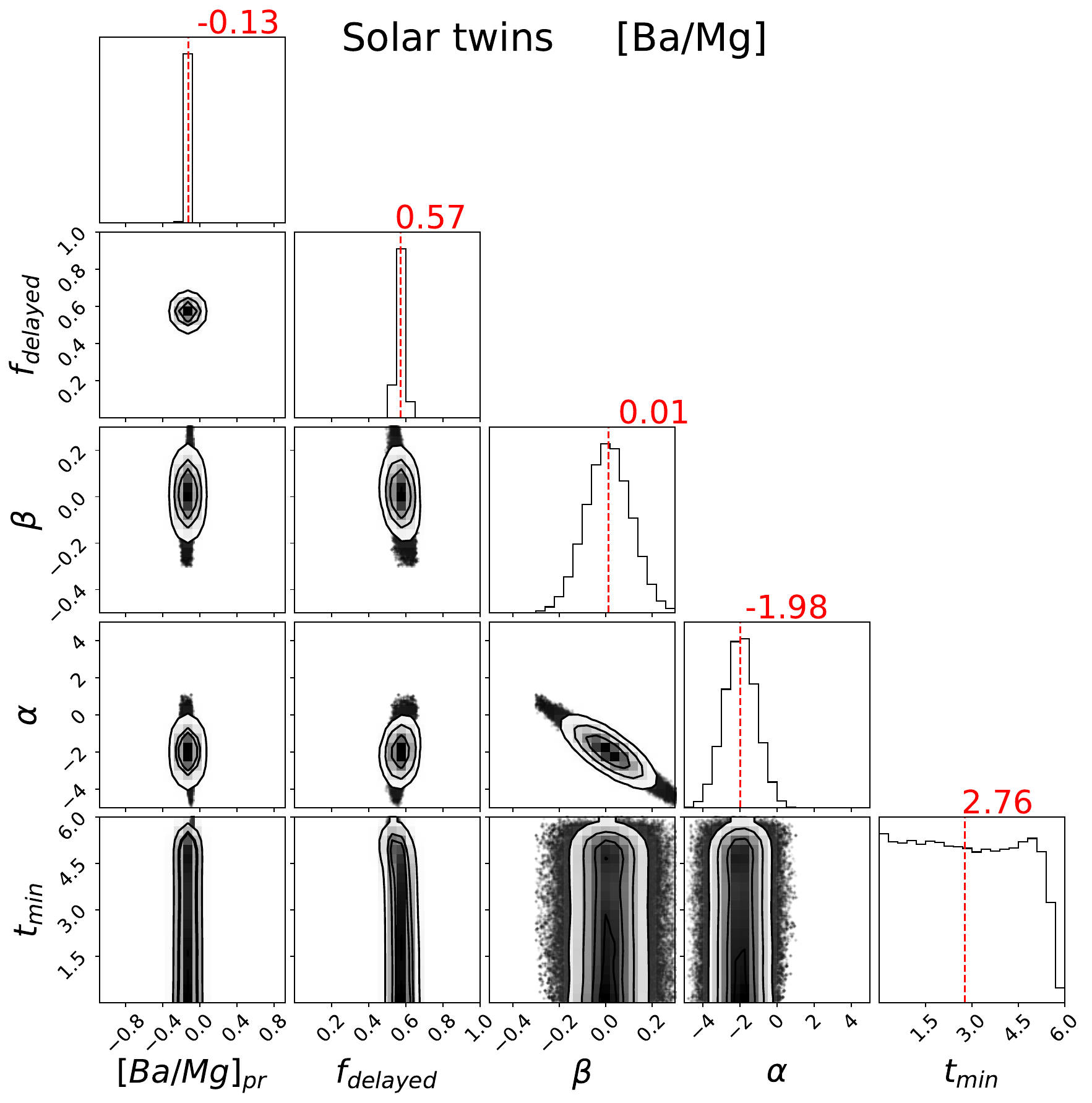}
	\end{minipage}
}
\caption{Posterior distributions for the [Ba/Mg] fit, for the LAMOST RC sample (left) and the solar twins (right). Diagonal panels show the 1D marginalized posteriors with their medians (red lines); off-diagonal panels show the 2D posteriors. The recovered DTD index is $\alpha \approx -1.2$ (LAMOST) and $\approx -1.8$ (twins), i.e.\ a rising DTD, somewhat shallower than that of Eu, with a delayed fraction $f_{\rm delayed}\approx 0.6$ in both samples.
\label{fig:corner_ba}}
\end{figure*}

\begin{figure*}[htb!]
\centering
\subfigure
{
	\begin{minipage}{0.46\linewidth}
	\centering
	\includegraphics[width=0.95\columnwidth]{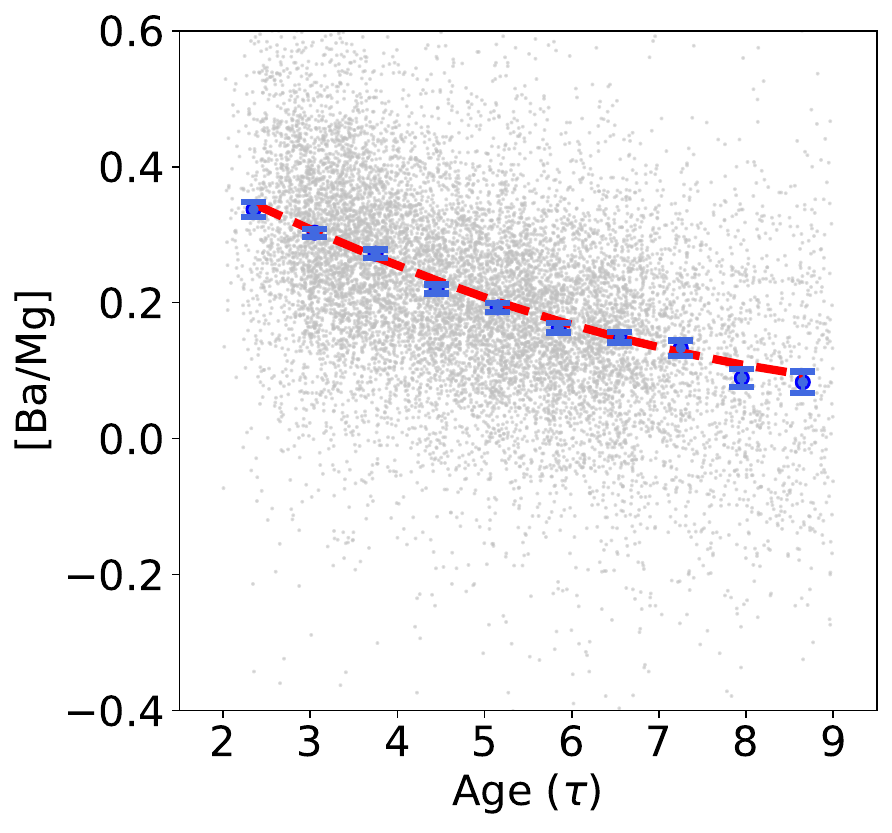}
	\end{minipage}
}
\subfigure
{
	\begin{minipage}{0.46\linewidth}
	\centering
	\includegraphics[width=0.95\columnwidth]{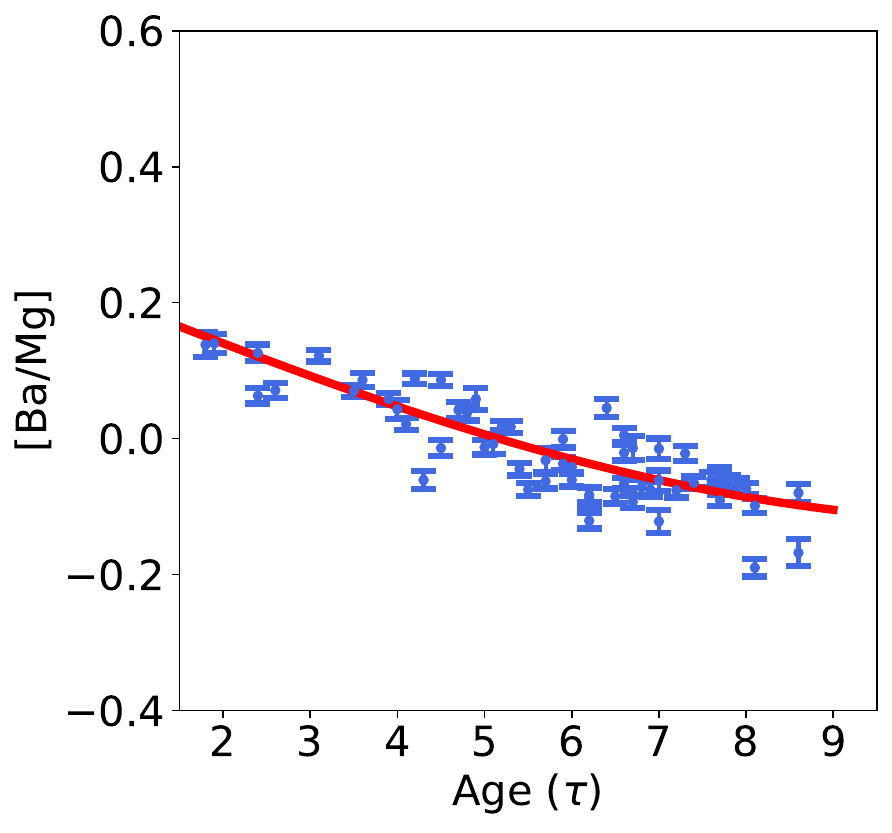}
	\end{minipage}
}
\caption{Chemical evolution model fits for [Ba/Mg]. \textit{Left}: The blue dots show the LAMOST median [Ba/Mg] versus stellar age, with error bars. The red dotted line shows the best-fit chemical model. The grey dots show the distribution of individual LAMOST sample stars. \textit{Right}: The blue dots show [Ba/Mg] for solar twins versus stellar age, with error bars.
\label{fig:results_ba}}
\end{figure*}

Barium is dominated at solar metallicity by the s-process in low- and intermediate-mass AGB stars, themselves a delayed channel. Fitting [Ba/Mg]($\tau$) (Figs.~\ref{fig:corner_ba} and \ref{fig:results_ba}, Table~\ref{tab:fits}), we again find a DTD that rises with delay time, $\alpha \approx -1.36$ (LAMOST) and $\approx -1.98$ (solar twins), i.e.\ somewhat shallower than for Eu, with a delayed fraction $f_{\rm delayed}({\rm today}) \approx 0.6$ in both samples. That the s-process element Ba shows a rising, strongly-delayed effective DTD is qualitatively expected: the AGB stars that produce it take $\gtrsim\!1$\,Gyr to evolve, so a substantial delayed contribution arises on independent grounds. We caution, however, that the steep rise of [Ba/Mg] toward young ages coincides with the long-standing ``barium puzzle,'' part of which has been attributed to a stellar-activity systematic in the strong Ba\,{\sc ii} lines rather than to nucleosynthesis; we return to this in Sect.~\ref{sec:discussion} and argue that Eu, a near-pure r-process element measured from weaker lines, is the cleaner of the two tracers.

\subsection{Summary of Fits}
\label{sec:summary}

Table~\ref{tab:fits} collects the best-fit parameters for all six fits. Three points stand out. First, the validation works: for Fe we recover a declining DTD ($\alpha>0$, ${\rm DTD}\propto t^{-1.2}$ to $t^{-1.6}$) and a high delayed fraction ($\approx\!0.8$), both in line with the known Type~Ia origin of iron. Second, Eu and Ba show the \emph{opposite} sign of $\alpha$: their DTDs \emph{rise} with delay time ($\alpha<0$), with $\approx\!60\%$ of each element in present-day stars coming from the delayed channel. Third, the two independent samples, LAMOST RC and solar twins, with very different abundance systematics, agree on the sign and rough magnitude of every parameter, which argues that the trends are not an artifact of either dataset.

Two parameters are common to all fits and only weakly constrained. The SFH rate is consistently small and slightly negative, $\beta \approx -0.1$ to $0.0$\,Gyr$^{-1}$; taken at face value this implies an e-folding time $|1/\beta|\gtrsim 10$\,Gyr, i.e.\ a star-formation rate that has been slowly declining, to within a factor of $\sim\!2$ constant, over the past $\sim\!10$\,Gyr, broadly consistent with solar-neighborhood SFH determinations \citep{Sysoliatina2021, Ruiz-Lara2020}. The minimum delay time $t_{\rm min}$ is poorly constrained in every fit (its posterior runs up against the prior; Fig.~\ref{fig:corner_fe}, Fig.~\ref{fig:corner_eu} and Fig.\ref{fig:corner_ba}); we therefore treat it as a nuisance parameter rather than a measurement.

\begin{deluxetable*}{llrrrrr}
\tablecaption{Best-fit chemical-evolution parameters for the six element--sample combinations. Values are posterior medians; the DTD follows ${\rm DTD}\propto\tau^{-\alpha}$ (Eq.~\ref{eq:DTDdef}), so $\alpha>0$ is a declining and $\alpha<0$ a rising DTD. Full posteriors are shown in Figs.~\ref{fig:corner_fe}, \ref{fig:corner_eu}, and \ref{fig:corner_ba}. \label{tab:fits}}
\tablehead{
\colhead{Element} & \colhead{Sample} & \colhead{$[{\rm X/Mg}]_{\rm pr}$} & \colhead{$f_{\rm delayed}$} & \colhead{$\beta$} & \colhead{$\alpha$} & \colhead{$t_{\rm min}$} \\
\colhead{} & \colhead{} & \colhead{(dex)} & \colhead{(today)} & \colhead{(Gyr$^{-1}$)} & \colhead{} & \colhead{(Gyr)}
}
\startdata
Fe & LAMOST RC   & $-0.49_{-0.29}^{+0.20}$ & $0.69_{-0.18}^{+0.15}$ & $-0.06_{-0.08}^{+0.08}$ & $+1.23_{-0.56}^{+0.50}$ & $0.69_{-0.35}^{+0.23}$ \\
Fe & Solar twins & $-0.66_{-0.24}^{+0.25}$ & $0.79_{-0.16}^{+0.09}$ & $-0.06_{-0.05}^{+0.05}$ & $+1.58_{-0.35}^{+0.30}$ & $0.78_{-0.28}^{+0.16}$ \\
Eu & LAMOST RC   & $-0.14_{-0.03}^{+0.02}$ & $0.60_{-0.05}^{+0.06}$ & $-0.10_{-0.10}^{+0.10}$ & $-2.58_{-1.61}^{+1.45}$ & $2.93_{-2.03}^{+2.07}$ \\
Eu & Solar twins & $-0.01_{-0.01}^{+0.01}$ & $0.40_{-0.02}^{+0.02}$ & $-0.02_{-0.10}^{+0.10}$ & $-2.05_{-0.96}^{+0.96}$ & $2.86_{-1.96}^{+2.05}$ \\
Ba & LAMOST RC   & $+0.07_{-0.03}^{+0.02}$ & $0.63_{-0.02}^{+0.02}$ & $-0.10_{-0.10}^{+0.10}$ & $-1.36_{-0.94}^{+0.93}$ & $2.88_{-1.93}^{+1.97}$ \\
Ba & Solar twins & $-0.13_{-0.02}^{+0.02}$ & $0.57_{-0.02}^{+0.02}$ & $0.01_{-0.10}^{+0.10}$ & $-1.98_{-0.87}^{+0.85}$ & $2.76_{-1.88}^{+1.95}$ \\
\enddata
\end{deluxetable*}

\section{Discussion}
\label{sec:discussion}

\subsection{Comparison with Previous Work}
Our finding of substantial delayed Eu enrichment ($f_{\rm delayed}\approx 0.6$) is broadly consistent with studies concluding that NSM are significant producers of r-process elements. However, we infer the \emph{shape} is qualitatively different: rather than a declining DTD, our age-based fit returns a DTD that \emph{rises} with delay time ($\alpha \approx -2.5$, ${\rm DTD}\propto\tau^{2.5}$ over $\sim\!1$--$10$\,Gyr). This is the opposite trend to the steep \emph{decreasing} power laws (${\rm DTD}\propto t^{-1.5}$ to $t^{-1.9}$) inferred from [Eu/Fe] versus [Fe/H] \citep{Cote2019, Simonetti2019}, and a striking enough result that we discuss possible reconciliations carefully before claiming a genuinely new channel.

Our result is most comparable to \citet{Tsujimoto2021}, who analyzed the same solar twins and likewise concluded that a substantial, long-delayed r-process contribution is required, with an NSM DTD index $n\approx0$--$0.5$ (i.e.\ flat to mildly rising with delay time) in addition to a prompt CCSNe site. Our inference agrees on the \emph{sign}, delayed enrichment continuing to accumulate to late times, but differs in two respects. First, \citet{Tsujimoto2021} worked with [r-process/Fe] and ascribed much of the age trend to radial migration (older twins born at smaller $R_{\rm GC}$, including bulge migrators); by referencing to Mg and testing against guiding radius (Sec.~\ref{sec:discussion}, Fig.~\ref{fig:rg}), we find a trend that persists across angular momentum and is therefore not readily explained by a birth-radius gradient alone. Second, our inferred effective DTD rises substantially more steeply ($\alpha\lesssim-2.0$, i.e.\ ${\rm DTD}\propto\tau^{2.0}$) than the flat-to-mild slope of \citet{Tsujimoto2021}; whether this difference reflects the different reference element, the much larger LAMOST sample, or a genuinely different inference is not clear.

We suggest that this difference reflects, at least in part, the fundamentally different observables being used. Studies using [Eu/Fe] versus [Fe/H] are sensitive to the \emph{absolute} DTD convolved with the full chemical evolution history, including the transition from thick to thin disk, the onset of Type Ia supernovae, and potential variations in the IMF or CCSNe yields. The ``knee'' in [Eu/Fe] at [Fe/H] $\sim -1.0$ can arise from multiple effects beyond the DTD shape. In contrast, our age-based approach at fixed solar metallicity directly measures the \emph{temporal derivative} of delayed enrichment accumulation, isolating the DTD signal from other evolutionary effects.

The apparent tension may also reflect the different delay-time ranges probed. Metallicity-based studies are most sensitive to enrichment at early times ([Fe/H] $\sim -2$ to $-1$, corresponding to $\sim$11--12\,Gyr ago), where the minimum delay time and the steep early rise of power-law DTDs dominate. Our age-based method probes enrichment over the past 10\,Gyr, where the long-delay tail of the DTD is most relevant. A DTD with both a steep early component and an extended late-time tail could potentially reconcile both sets of constraints.

\subsection{Implications for Neutron Star Merger Physics}

The minimum delay time is only weakly constrained ($t_{\rm min}\sim 3$\,Gyr, but with a posterior that runs up against the prior), so we do not over-interpret it. Forming a merging double neutron star requires (1) two massive stars in a close binary, (2) two supernova explosions that leave the binary bound, and (3) orbital decay via gravitational-wave radiation. Theoretical models predict minimum merger times of $\sim\!10$--$100$\,Myr for the closest binaries, with a tail extending to $\gtrsim\!1$\,Gyr \citep{Dominik2012}. The more surprising aspect of our result is not $t_{\rm min}$ but the \emph{rise} of the inferred DTD with delay time: a standard NSM population produces a declining DTD, set by the $\propto a^{4}$ scaling of gravitational-wave inspiral time with initial separation. A DTD that increases with delay time is therefore difficult to produce with NS--NS mergers alone, and, if the result survives the systematic tests below, would point to an additional, more strongly delayed r-process channel.

\subsection{Advantages and Limitations of the Age-Based Approach}

The primary advantage of using stellar ages is directness: time is directly observable, whereas inferring temporal evolution from [Fe/H] requires modeling the full metallicity enrichment history. By working at fixed solar metallicity, we minimize sensitivity to metallicity-dependent yields, radial migration effects, and the thick-versus-thin disk transition. The differential measurement of [Eu/Mg] isolates delayed enrichment against the purely prompt Mg background.

However, our approach has several limitations. First, the strong covariance $\alpha$--$\beta$ means we constrain primarily the \emph{relative} accumulation rate of delayed enrichment versus the SFH, not the DTD index and the SFH rate fully independently; breaking this requires an external SFH constraint (Sect.~\ref{sec:future}). Second, the solar-twin sample is small ($79$ stars) and the LAMOST ages, while plentiful ($>\!10{,}000$ stars), are individually uncertain at the $\sim\!30$--$40\%$ level (and mildly compress the age scale toward the training mean), so we rely on binned medians; larger samples with smaller age errors, spanning a broader age range (ideally $0$--$12$\,Gyr), will sharpen all parameters, particularly $t_{\rm min}$ and the DTD shape. Third, we adopt a power-law DTD and an exponential SFH for tractability; more flexible forms (SFH bursts, or multiple DTD components) may be warranted as the data improve, and could be important if the rising-DTD signal is in fact the superposition of a declining NSM component and a second, more delayed channel.

A systematic that deserves particular attention is radial migration \citep{Tsujimoto2021}. Our cut to [Fe/H] $\approx 0$ selects stars that reached solar metallicity, but stars of different ages did so via different evolutionary paths and, after migration, may have been born at different Galactocentric radii with different local enrichment histories. Because solar metallicity is reached earlier at smaller radii, an age sequence at [Fe/H]$\,\approx 0$ is partly a sequence in birth radius, so a radial gradient in birth [X/Mg] could imprint an age-dependent trend that mimics or dilutes the DTD signal.

Our use of Mg as the reference element mitigates this effect, though it may not fully remove it. In using $[{\rm Eu/Mg}] = [{\rm Eu/Fe}] - [{\rm Mg/Fe}]$, i.e. referencing to Mg rather than Fe, we remove by construction the $[{\rm Mg/Fe}]$ ($\alpha$-clock) term, which carries most of the known radial and age structure of the $[{\rm X/Fe}]$ ratios and is driven by the varying Fe (Type~Ia) contribution; this is the same Fe-driven structure that dominates the [r-process/Fe] trends used by \citet{Tsujimoto2021}. What survives is the radial gradient of $[{\rm Eu/Mg}]$ itself, the r-process-to-$\alpha$ ratio, which may be covariant with the temporal signal and which Mg cannot cancel.

\begin{figure*}[htb!]
\centering
\subfigure
{
	\begin{minipage}{0.46\linewidth}
	\centering
	\includegraphics[width=0.95\columnwidth]{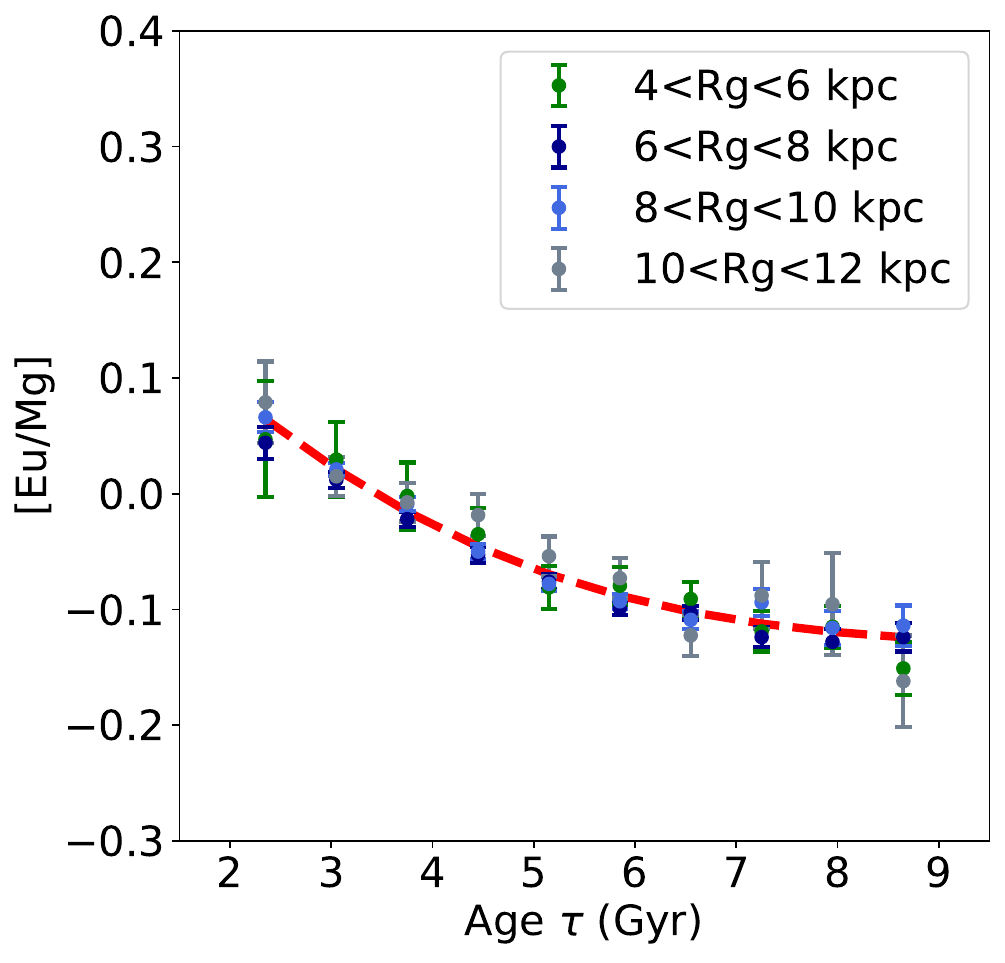}
	\end{minipage}
}
\subfigure
{
	\begin{minipage}{0.46\linewidth}
	\centering
	\includegraphics[width=0.95\columnwidth]{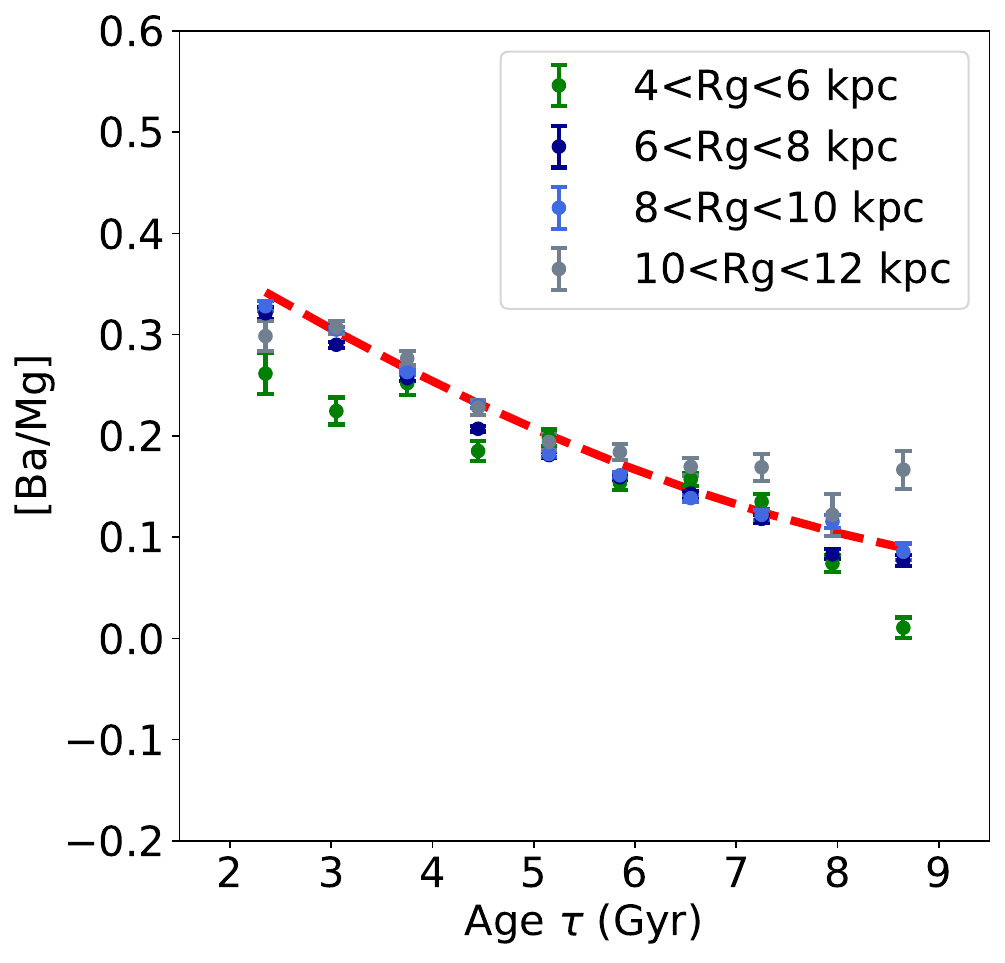}
	\end{minipage}
}
\caption{\textit{Left}: The dots with error bars show the LAMOST median [Eu/Mg] versus stellar age for stars of different guiding radii. The red dotted line shows the best-fit chemical model. \textit{Right}: The dots with error bars show [Ba/Mg] versus stellar age.
\label{fig:rg}}
\end{figure*}

To bound this residual directly, we split the LAMOST sample by guiding-center radius (a proxy for present orbital angular momentum) and re-examine the abundance--age trends. The [Eu/Mg] and [Ba/Mg] ratios for stars of different guiding radii are shown in Fig.~\ref{fig:rg}: both exhibit the same negative correlation with stellar age across the range of guiding radii probed. Although these stars are on orbits of different angular momentum, they display closely analogous [Eu/Mg]$(\tau)$ and [Ba/Mg]$(\tau)$ trends, indicating that the signal is not a simple radial gradient aliased onto age and is therefore unlikely to be dominated by radial migration. The test is necessarily imperfect, churning scrambles birth radius relative to present guiding radius, but a trend that is stable across guiding radius is the strongest available constraint on the migration contribution.

A second systematic, specific to barium, is the ``barium puzzle.'' The steep rise of [Ba/Mg] toward young ages that we recover (Fig.~\ref{fig:results_ba}) echoes the long-standing finding that [Ba/Fe] increases sharply with decreasing age in open clusters and young field stars, reaching $\sim\!+0.6$\,dex below $100$\,Myr \citep{DOrazi2009}, a rise that standard AGB s-process yields in chemical-evolution models cannot reproduce \citep{Maiorca2012}. Crucially, this overabundance is not accompanied by a corresponding rise in the other s-process elements La--Sm, and while an intermediate (i-) process origin has been proposed \citep{Mishenina2015}, \citet{ReddyLambert2017} and \citet{Spina2020} have argued that it is largely a \emph{spectroscopic} effect: the strong Ba\,{\sc ii} lines form high in the photosphere, where the enhanced microturbulence of young, chromospherically active stars inflates the inferred Ba abundance under standard 1D-LTE analysis. If part of the young-star Ba rise is such an activity-driven systematic, it is a concrete example of the age-correlated abundance error that would, in principle, also affect Eu. Two considerations limit the concern for our central result. First, Eu is a near-pure r-process element measured from weaker lines that are far less sensitive to the activity/microturbulence effect implicated in the Ba puzzle, so the cleaner Eu trend is the more robust of the two. Second, the barium puzzle is most acute below $\sim\!200$\,Myr, whereas our RC sample is cut at $\tau>2$\,Gyr; the solar twins do reach $0.7$\,Gyr and so overlap the affected regime, which is one more reason to weight the LAMOST RC result and to treat the steep Ba slope with more caution than the Eu slope.
Finally, our analysis assumes chemical equilibrium (instantaneous adjustment), which requires $\tau_{\rm adj} \ll \tau_{\rm evol}$. This is well satisfied at solar metallicity ($\tau_{\rm adj} \sim 1$\,Gyr vs.\ the SFH e-folding time $1/|\beta| \sim 10$\,Gyr) but may break down at lower metallicities where gas masses are smaller, in rapidly quenching systems, or during starburst episodes where $\tau_{\rm evol}$ becomes comparable to $\tau_{\rm adj}$.

\subsection{Future Directions}
\label{sec:future}

Several avenues can strengthen age-based DTD constraints:

\textbf{Larger samples:} Combining solar-twin samples \citep{Spina2018, Bedell2018} with asteroseismic surveys (TESS, Kepler) and large spectroscopic programs (e.g.\ LAMOST) could provide thousands of stars with ages and [Eu/Mg] measurements. This would tighten all parameters, particularly the minimum delay time and any deviations from a simple power-law DTD.

\textbf{Independent SFH constraints:} Detailed modeling of the Gaia color-magnitude diagram can provide an independent constraint on $\beta$, breaking the $\alpha$--$\beta$ degeneracy \citep{Ruiz-Lara2020}. Pinning $\beta$ externally would directly sharpen the inferred DTD index $\alpha$, which is the quantity of greatest physical interest here.

\textbf{Additional elements:} Extending the analysis to other neutron-capture elements (La, Nd, Sm, Y, and Sr), with their different r-/s-process mixes, would test whether the rising-DTD behavior tracks the r-process fraction and helps isolate the channel responsible.

\textbf{Spatial variations:} Applying the method across different Galactocentric radii could reveal spatial variations in the DTD or SFH, probing inside-out disk growth and radial migration effects.

\textbf{Non-parametric DTDs:} With sufficiently large samples, one could fit the DTD non-parametrically (e.g., in age bins) rather than assuming exponential forms, providing more flexible constraints on the true DTD shape.

\section{Conclusions}
\label{sec:conclusions}

We have developed and applied a new, age-based approach to constraining the delay time distribution of r- and s-process sources for the thin disk stars, using [X/Mg] versus stellar age at fixed solar metallicity. Working differentially against Mg, and in the regime where chemical-equilibrium models are expected to apply, creates a very direct connection between the DTD shape and the abundance--age trend, with minimal recourse to a full metallicity-evolution model. Our principal findings are:

\begin{enumerate}
\item \textbf{We succeeded in validating this approach by deriving the DTD for iron.} Applied to [Fe/Mg] in our data set, it recovers a declining DTD (${\rm DTD}\propto t^{-1.2}$ to $t^{-1.6}$) and a delayed fraction $\approx\!70\%$, both in good agreement with the well-established Type~Ia origin of iron.
\item \textbf{We find that the majority of the Eu and Ba to date comes from a channel that appears  substantially delayed.} For more than $10{,}000$ LAMOST RC thin disk stars (cross-checked with solar twins), $\approx\!60\%$ of both the Eu and the Ba in stars forming today originate from delayed ($\gtrsim\!1$\,Gyr) channels.
\item \textbf{These effective DTDs actually appear to rise with delay time.} In contrast to iron, and to the declining DTDs that had been inferred from [Eu/Fe] versus [Fe/H], the Eu and Ba effective DTDs we infer \emph{increase} with delay time ($\alpha\approx-2.6$ and $-1.4$, respectively). Because all abundances are referenced to Mg, and because the same trend is recovered across stars of different evolutionary stages (main sequence and giants) and guiding-center radii, the effect is probably not attributable to stellar photosphere physics of Galactic radial migration. The Eu result is the more robust of the two, as the steep [Ba/Mg] rise might be partly affected by the activity-related ``barium puzzle'' systematic in young ($<$1 Gyr) stars. If the rising trend survives further tests for abundance systematics, it points to a delayed neutron-capture channel, with greater delays than canonical NS--NS mergers.
\end{enumerate}

As asteroseismic surveys and large spectroscopic programs deliver better ages and neutron-capture abundances for extensive stellar samples, this age-based approach could yield increasingly precise constraints on the sites and timescales of r- and s-process nucleosynthesis, complementing gravitational-wave observations of neutron-star mergers and abundance patterns in metal-poor stars.

\section*{Acknowledgments}
We thank the referee for the suggestions that have improved the clarity of the manuscript.
We used data from the Guoshoujing Telescope (LAMOST). LAMOST is a National Major Scientific Project built by the Chinese Academy of Sciences. Funding for the project has been provided by the National Development and
Reform Commission. LAMOST is operated and managed by the National Astronomical Observatories, Chinese Academy of Sciences. Its website is https://www.lamost.org.

We also used data from the European Space Agency mission Gaia (http://www.cosmos.esa.int/gaia), processed by the Gaia Data Processing and Analysis Consortium (DPAC; see http://www.cosmos.esa.int/web/gaia/dpac/consortium). 

This work used the \texttt{emcee} \citep{Foreman-Mackey2013} and XGBoost \citep{Chen2016} packages, and used \emph{Claude.ai} for language checking and editing. 
\section*{Data Availability}
The stellar parameters and abundances underlying this work are public: the LAMOST DR9 value-added catalog \citep{ZhangM2025}, and the solar-twin abundances of \citet{Spina2018} and \citet{Bedell2018}. The derived spectroscopic ages, the binned [X/Mg]($\tau$) measurements, and the MCMC posteriors will be made available in electronic form in Zenodo, doi:10.5281/zenodo.22113880, in FITS format.

We acknowledge financial support from the National Natural Science Foundation of China (NSFC; Grant No.12303025)

\appendix

\section{Detailed Derivation of Equilibrium Abundances}

\subsection{Mass Balance Equations}

The Eu mass from delayed sources in the ISM evolves according to:
\begin{equation}
\frac{dM_{\rm Eu}^{\rm delay}}{dt} = \text{(Production)} - \text{(Loss to SF)} - \text{(Loss to outflows)} + \text{(Recycling)}
\end{equation}

\textbf{Production term:} Stars formed at rate $\psi(t')$ at time $t'$ produce delayed Eu at the current time $t$ at a rate:
\begin{equation}
\frac{dM_{\rm Eu}^{\rm delay}}{dt}\bigg|_{\rm prod} = y_{\rm Eu}^{\rm delay} \int_{0}^{t-t_{\rm min}} \psi(t') \cdot {\rm DTD}(t-t') \, dt'
\end{equation}

\textbf{Units:} [dimensionless] $\times$ [M$_\odot$/Gyr] $\times$ [Gyr$^{-1}$] $\times$ [Gyr] = [M$_{\odot}$/Gyr]

\textbf{Loss to star formation:}
\begin{equation}
\frac{dM_{\rm Eu}^{\rm delay}}{dt}\bigg|_{\rm SF} = -Z_{\rm Eu}^{\rm delay} \cdot \psi(t)
\end{equation}

where $Z_{\rm Eu}^{\rm delay} \equiv M_{\rm Eu}^{\rm delay}/M_g$ [dimensionless] is the Eu mass fraction.

\textbf{Loss to outflows:} With mass loading factor $\eta \equiv \dot{M}_{\rm out}/\dot{M}_*$:
\begin{equation}
\frac{dM_{\rm Eu}^{\rm delay}}{dt}\bigg|_{\rm out} = -Z_{\rm Eu}^{\rm delay} \cdot \eta \cdot \psi(t)
\end{equation}

Following \citet{Weinberg2017}, $\eta \approx 2$ for the Milky Way disk.

\textbf{Recycling:} A fraction $r$ of stellar mass returns to the ISM:
\begin{equation}
\frac{dM_{\rm Eu}^{\rm delay}}{dt}\bigg|_{\rm rec} = +Z_{\rm Eu}^{\rm delay} \cdot r \cdot \psi(t)
\end{equation}

For a \citet{Kroupa2001} IMF, $r \approx 0.4$ \citep{Weinberg2017}.

\subsection{Equilibrium Solution}

Under chemical equilibrium (instantaneous adjustment), setting $dM_{\rm Eu}^{\rm delay}/dt = 0$ at each time $t$:
\begin{equation}
y_{\rm Eu}^{\rm delay} \int \psi(t') \cdot {\rm DTD}(t-t') dt' = Z_{\rm Eu}^{\rm delay} \cdot (1 + \eta - r) \cdot \psi(t)
\end{equation}

\textbf{Important:} This does not mean $Z_{\rm Eu}^{\rm delay}(t)$ is constant in time. Rather, at each instant $t$, the abundance adjusts to balance the \emph{current} production integral (which grows as delayed sources accumulate) against the \emph{current} loss rate $\psi(t)$. As $t$ increases, the production integral increases, so $Z_{\rm Eu}^{\rm delay}(t)$ increases, even though the system remains in instantaneous equilibrium at each moment.

Solving for the abundance:
\begin{equation}
Z_{\rm Eu}^{\rm delay}(t) = \frac{y_{\rm Eu}^{\rm delay}}{1 + \eta - r} \int_{t_{\rm min}}^{\min(t,t_{\rm max})} \frac{\psi(t')}{\psi(t)} \cdot {\rm DTD}(t-t') \, dt'
\end{equation}

Similarly, for prompt Eu:
\begin{equation}
Z_{\rm Eu}^{\rm pr} = \frac{y_{\rm Eu}^{\rm pr}}{1 + \eta - r}
\end{equation}

And for Mg:
\begin{equation}
Z_{\rm Mg} = \frac{y_{\rm Mg}^{\rm cc}}{1 + \eta - r}
\end{equation}

\subsection{Cancellation in Ratios}

When forming ratios, the factor $(1 + \eta - r)$ cancels:
\begin{equation}
\frac{Z_{\rm Eu}^{\rm pr}}{Z_{\rm Mg}} = \frac{y_{\rm Eu}^{\rm pr}}{y_{\rm Mg}^{\rm cc}}, \quad \frac{Z_{\rm Eu}^{\rm delay}}{Z_{\rm Mg}} = \frac{y_{\rm Eu}^{\rm delay}}{y_{\rm Mg}^{\rm cc}} \mathcal{R}(t)
\end{equation}

where we define:
\begin{equation}
\mathcal{R}(t) \equiv \int_{t_{\rm min}}^{\min(t,t_{\rm max})} \frac{\psi(t')}{\psi(t)} \cdot {\rm DTD}(t-t') \, dt'
\end{equation}

For the Milky Way with $\eta \approx 2$ and $r \approx 0.4$, we have $(1 + \eta - r) = 2.6$, but this cancels in all observable abundance ratios.

\subsection{Evaluation of $\mathcal{R}(t)$}

For the exponential SFR $\psi(t) = \psi_0 e^{\beta t}$ and the power-law DTD$(\tau) = A\,\tau^{-\alpha}$ (Eq.~\ref{eq:DTDdef}), changing variables to the delay $\tau = t - t'$ gives
\begin{equation}
\mathcal{R}(t) = A \int_{t_{\rm min}}^{t_{\rm eff}} e^{-\beta\tau}\,\tau^{-\alpha}\, d\tau ,
\label{eq:Rint}
\end{equation}
with $t_{\rm eff} = \min(t, t_{\rm max})$. For a power-law DTD this integral does not reduce to elementary functions for general $\alpha$; it is a (generalized) incomplete-gamma integral,
\begin{equation}
\int_{t_{\rm min}}^{t_{\rm eff}} e^{-\beta\tau}\,\tau^{-\alpha}\, d\tau
= \beta^{\alpha-1}\left[\Gamma\!\left(1-\alpha,\,\beta t_{\rm min}\right) - \Gamma\!\left(1-\alpha,\,\beta t_{\rm eff}\right)\right]
\quad (\beta>0),
\end{equation}
where $\Gamma(1-\alpha, \beta_{t_{\rm min}})$ and $\Gamma(1-\alpha, \beta_{t_{\rm eff}})$ are the upper incomplete gamma functions. In practice, we evaluate Eq.~(\ref{eq:Rint}) numerically inside the likelihood, which also handles $\beta\le 0$ and arbitrary $\alpha$ without special cases. The DTD normalization $A$ is fixed by $\int_{t_{\rm min}}^{t_{\rm max}} {\rm DTD}(\tau)\, d\tau = 1$, and cancels when $\mathcal{R}(t)$ is divided by $\mathcal{R}(t_{\rm now})$ in the observable model below.

\subsection{Complete Observable Model}

The full model for [Eu/Mg] as a function of stellar age, expressed in terms of the physically meaningful parameter $f_{\rm delayed}({\rm today})$, is:
\begin{equation}
\left[\frac{\rm Eu}{\rm Mg}\right](\tau_{\rm age}) = \left[\frac{\rm Eu}{\rm Mg}\right]_{\rm pr} + \log_{10}\left[1 + \frac{f_{\rm delayed}({\rm today})}{1 - f_{\rm delayed}({\rm today})} \cdot \frac{\mathcal{R}(t_{\rm now} - \tau_{\rm age})}{\mathcal{R}(t_{\rm now})}\right]
\end{equation}

where the conversion between $f$ and $f_{\rm delayed}({\rm today})$ is:
\begin{equation}
f = \frac{f_{\rm delayed}({\rm today})}{\mathcal{R}(t_{\rm now}) \cdot [1 - f_{\rm delayed}({\rm today})]}
\end{equation}

\bibliographystyle{aasjournal}
\bibliography{main.June21.2026.bib}
\end{document}